\documentclass[12pt]{spieman}  % 12pt font required by SPIE;
\usepackage{amsmath,amsfonts,amssymb}
\usepackage{graphicx}
\usepackage{setspace}
\usepackage{tocloft}
\usepackage{physics}

\usepackage{xcolor}
\usepackage{lineno}

\newcommand{\red}[1]{{\color{black}#1}}
\newcommand{\comment}[1]{}

\title{Pyxis: A ground-based demonsstrator for formation-flying optical interferometry}

\author[a,*]{Jonah T. Hansen}
\author[b]{Samuel Wade}
\author[a]{Michael J. Ireland}
\author[b]{Tony D. Travouillon}
\author[a]{Tiphaine Lagadec}
\author[b]{Nicholas Herrald}
\author[b]{Joice Mathew}
\author[a,c]{Stephanie Monty}
\author[a,d]{Adam D. Rains}
\affil[a]{Research School of Astronomy and Astrophysics, College of Science, Australian National University, ACT 2611, Australia}
\affil[b]{Advanced Instrumentation Technology Centre, Research School of Astronomy and Astrophysics, College of Science, Australian National University, ACT 2611, Australia}
\affil[c]{Institute of Astronomy, University of Cambridge, Madingley Rd, Cambridge, CB3 0HA, UK}
\affil[d]{Division of Astronomy and Space Physics, Department of Physics and Astronomy, Uppsala University, Box 516, 75120 Uppsala, Sweden}

\cftpagenumbersoff{figure}
\cftpagenumbersoff{table}

\begin{document} 
\maketitle

\begin{abstract}
In the past few years, there has been a resurgence in studies towards space-based optical/infrared interferometry, particularly with the vision to use the technique to discover and characterise temperate Earth-like exoplanets around solar analogues. One of the key technological leaps needed to make such a mission feasible is demonstrating that formation flying precision at the level needed for interferometry is possible. Here, we present \textit{Pyxis}, a ground-based demonstrator for a future small satellite mission with the aim to demonstrate the precision metrology needed for space-based interferometry. We describe the science potential of such a ground-based instrument, and detail the various subsystems: three six-axis robots, a multi-stage metrology system, an integrated optics beam combiner and the control systems required for the necessary precision and stability. We end by looking towards the next stage of \textit{Pyxis}: a collection of small satellites in Earth orbit.
\end{abstract}

% Include a list of up to six keywords after the abstract

% Include email contact information for corresponding author
{\noindent \footnotesize\textbf{*}Email:  \linkable{jonah.hansen@anu.edu.au} }

%\begin{spacing}{2}   % use double spacing for rest of manuscript

\section{Introduction and Background}
\label{sec:pyxis_intro}  % \label{} allows reference to this section

High angular resolution astrophysics, in particular optical/infrared (IR) interferometry, is in a golden age. Instruments such as GRAVITY \cite{Abuter2017} and MATISSE \cite{Lopez2022} at the VLTI and MIRC-X \cite{Anugu2020} on the CHARA array have produced stunning scientific results, including imaging the starspots on a distant giant star\cite{Roettenbacher2016}, the first astrometric confirmation of a planet\cite{Lacour2021}, and of course, the characterisation of a supermassive compact object at the centre of our galaxy\cite{Gravity2019}, \red{contributing} to the 2020 Nobel Prize in physics. However, there are still questions yet unexplored that only interferometry will be able to probe.

One such avenue is the direct imaging of exoplanets. One of the major goals of exoplanet research is identifying potentially habitable worlds that may harbour life, and to that end one needs access to the atmosphere of the planet to look for biosignatures \cite{Schwieterman2018}. Transmission spectroscopy is promising for the characterisation of hydrogen rich atmospheres, but is challenging for terrestrial atmospheres\cite{DiamondLowe2020}. Hence direct imaging is one of the only techniques available to obtain atmospheric spectra from terrestrial planets. In order to accomplish this, however, we require minimising the contrast between a planet and its host star---which for terrestrial planets lies in the mid-infrared (MIR). To obtain the sensitivity needed, as well as avoiding telluric contamination, we also require these telescopes to be in space, producing challenges in having large coronagraphic apertures with the required angular resolution.

Hence space-based interferometry has long been recognised as \red{an ideal and cost-effective} way to take MIR spectra of Earth-like planets\cite{Cockell2009,Defrere2018}, and is simulated to have similar or greater yield compared to the largest launchable $>$\$10B coronagraphic telescopes in detecting any habitable planet biosignatures \cite{Kammerer2018,LIFE1}. Once the required aperture diameter becomes too large to construct mechanically or to launch, the only option is to launch parts of the mirror as separate light collector spacecraft with light combined in a beam combiner spacecraft—becoming a space interferometer. 

Many previous studies have been made of space interferometer missions, in one of two categories. These were either connected element interferometers\cite{Leisawitz2007,Unwin2008} which \red{are limited in the maximum baselines they can achieve}, or formation flying interferometers \red{usually situated at the Sun-Earth L2 point\cite{LeDuigou2006,Cockell2009}}. In the late 2000s, both NASA and ESA shelved plans for large scale space-based interferometers (TPF-I\cite{Beichman1999} and \textit{Darwin}\cite{Cockell2009} respectively), primarily due to a lack of understanding of the potential planet yield of such a mission, as well as technical unreadiness. Since that time, however, missions such as \textit{Kepler} \cite{Borucki2010} and more recent exoplanet missions (e.g. TESS\cite{Ricker2015} and CHEOPS \cite{Broeg2013}) have greatly increased our knowledge of planet demographics to the point where simulations of a \red{Flagship-class large MIR space interferometer, based on the point designs of TPF-I/\textit{Darwin},} have shown it would detect approximately 20 Earth-like planets \cite{LIFE1,LIFE2}. This renewal of interest in space interferometry has led to the development of the Large Interferometer For Exoplanets (LIFE) initiative, a revival of the TPF-I/\textit{Darwin} concept to detect and characterise planets in the MIR \cite{LIFE1}.  

In late 2021, ESA released its Voyage 2050 plan, for which the characterisation of planets in the MIR was one of the top priorities for a large \red{L-class} mission \cite{ESAVoyage2021}. As a caveat to this recommendation, however, was the requirement to prove that such a space interferometer mission would be feasible technologically, as well as scientifically. Two critical technology areas have not been at an adequate level to progress the formation flying optical and infrared interferometry missions: compact, cryogenic compatible nulling beam combiners (a target of the Nulling Interferometer Cryogenic Experiment (NICE)  \cite{Ranganathan2022}), and formation flying itself, including metrology systems. It is this second technology that is the primary purpose of \textit{Pyxis}, the subject of this paper, though other investigations into formation flying interferometry are currently ongoing \cite{Dandumont2020,Matsuo2022}.

\begin{figure}
    \centering
    \includegraphics[width=\columnwidth]{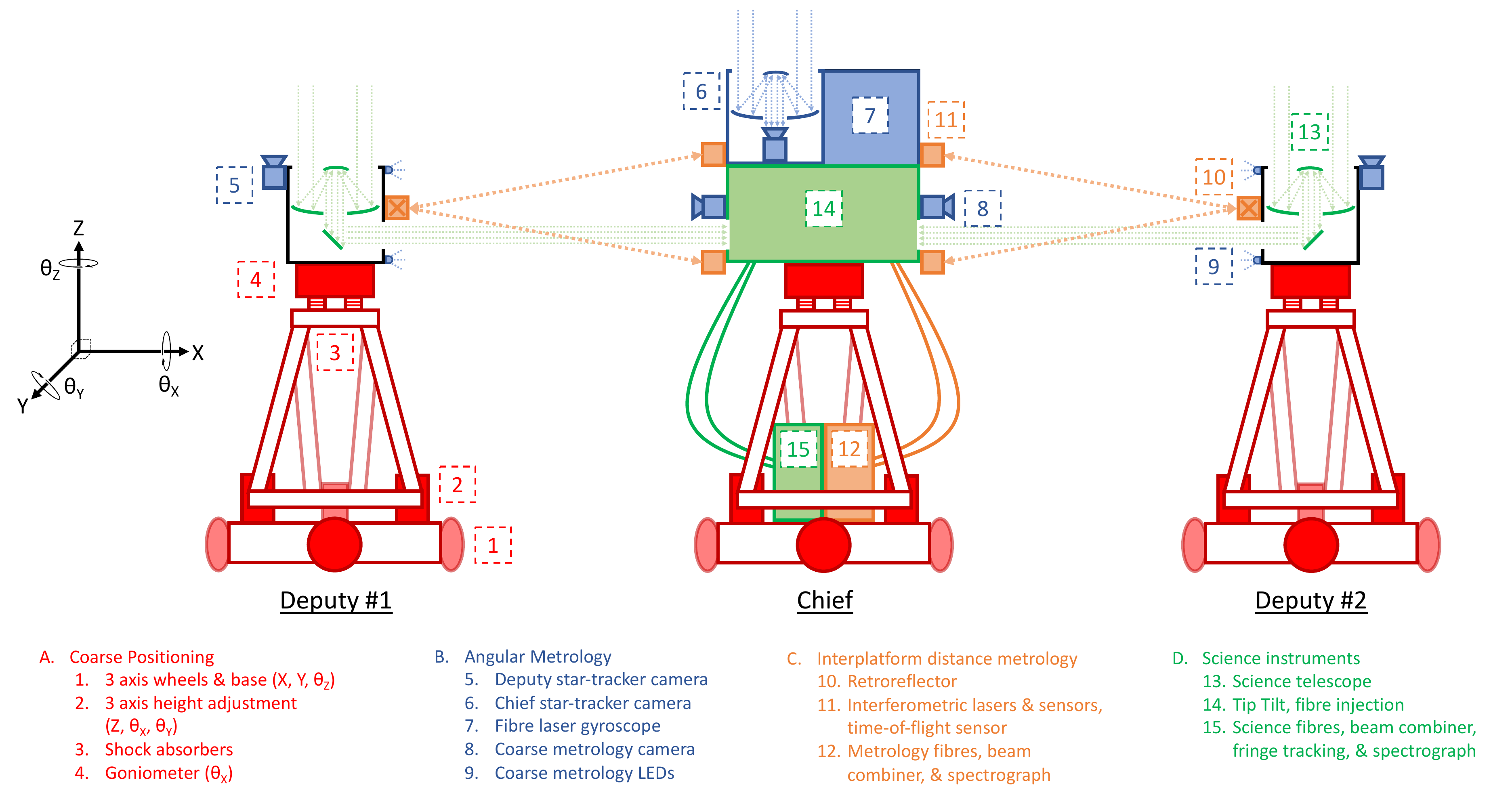}
    \vspace{0.3cm}
    \caption{Schematic showing \textit{Pyxis}' two science telescope platforms (known as the deputies) and the single central beam combining platform (known as the chief), each of which are separate wheeled 6-axis platforms able to move to position on a moderately flat surface and then track spatial and angular positions. Starlight is focused and collimated by the telescope primary and secondary mirrors, before reflecting off of 45$^{\circ}$ flat mirrors to the tip/tilt, fibre injection, and beam combiner systems on the central platform. Spatial locations are measured using accelerometers and a white-light fast metrology system, with the frame of reference of the central beam combiner determined by a combination of a fibre optic gyroscope and star tracker. In addition to the electro-mechanical and control systems, the \textit{Pyxis} concept can be broadly broken into the following four systems: \textbf{Coarse Positioning (Red):} \textbf{1}: 3 axis wheels \& base (X, Y, $\theta_{\rm Z}$); \textbf{2}: 3 axis height adjustment (Z, $\theta_{\rm X}$, $\theta_{\rm Y}$); \textbf{3}: Shock absorbers; \textbf{4}: Goniometer ($\theta_{\rm X}$). \textbf{Angular Metrology (Blue):} \textbf{5}: Deputy star-tracker camera; \textbf{6}: Chief star-tracker camera; \textbf{7}: Fibre optic gyroscope; \textbf{8}: Coarse metrology camera; \textbf{9}: Coarse metrology LEDs. \textbf{Interplatform distance metrology (Orange):} \textbf{10}: Retro-reflector; \textbf{11}: Interferometric lasers \& sensors, time-of-flight sensor; \textbf{12}: Metrology fibres, beam combiner, \& spectrograph; \textbf{Science instruments (Green):} \textbf{13}: Science telescope; \textbf{14}: Tip tilt, fibre injection; \textbf{15}: Science fibres, beam combiner, fringe tracking, \& spectrograph.}
    \label{fig:pyxis_schematic}
\end{figure}

\textit{Pyxis} is a multi-platform, linear-formation, robotic ground-based optical interferometer in development at the Australian National University's (ANU) Research School of Astronomy and Astrophysics (RSAA) located at Mt Stromlo Observatory. It will serve as a crucial technology demonstration of formation control and metrology for future formation flying space-interferometry missions and enable far more flexible ground-based stellar interferometry. A schematic of the \textit{Pyxis} interferometer is found in Figure \ref{fig:pyxis_schematic}, highlighting the major components and subsystems. Here we distinguish between the side collector platforms as ``deputies'', and the central beam combining platform as the ``chief''.

\textit{Pyxis} has a number of novel key features that will allow it to achieve its goal of formation flying interferometry. Firstly, it utilises a frame of reference tied to a precision star tracker on a movable platform, rather than the Earth itself. This is made possible through the use of newly affordable MEMS (microelectromechanical system) accelerometers and a fibre laser gyroscope to define this frame of reference, and will be discussed more in Section \ref{sec:control}. Secondly, we implement a multi-stage metrology system, using camera-based coarse metrology supplemented by a time of flight (TOF) sensor; and a sub-wavelength path-differential interferometric metrology sensor using laser diodes. These systems allow us to vastly simplify our system architecture and will be discussed in Section \ref{sec:metrology}. Finally, \textit{Pyxis} has a linear array architecture that can be used in Low Earth Orbit (LEO)~\cite{Hansen2020} and has a continuous range of reconfigurable baselines, nominally between 1 and 60~m.

This paper is structured as follows: the remainder of Section \ref{sec:pyxis_intro} details the scientific potential and aims of \textit{Pyxis}; Section \ref{sec:mechanical} describes the mechanical interface and architecture of the interferometer; Section \ref{sec:metrology} details the metrology system; Section \ref{sec:BC} discusses the beam combiner; and finally Section \ref{sec:control} will describe the control systems. 

\subsection{Scientific Aims}
\label{sec:science}

While the primary purpose of \textit{Pyxis} is to act as a demonstrator for formation-flying interferometry, it will be well placed to make important, unique astrophysical measurements on its own. \textit{Pyxis} is designed to nominally work in the $R$ band with a wavelength range between $\sim$620 and 760~nm, where the upper bound at 760~nm is due to the atmospheric telluric band corresponding to the Fraunhofer A O$_2$ band and the lower bound corresponding to the single-mode cutoff of our 630~nm fibres. The science telescopes on each deputy platform (see Figure \ref{fig:pyxis_schematic}) have an aperture diameter of 94~mm, and are expected to achieve a 10\% throughput including fibre coupling. We thus expect to be able to achieve a limiting magnitude of approximately R$\sim$6 with 5~ms exposures, corresponding to approximately 10 pixels per spectral channel. The aperture size was chosen due to the only moderate sensitivity gains when increasing the aperture beyond Fried's parameter ($r_0 = 5-10$~cm) without adaptive optics. Instead, the sensitivity of \textit{Pyxis} is increased through having a simple optical design (covered in Section \ref{sec:BC}). 

\begin{figure}
    \centering
    \includegraphics[width=0.9\columnwidth]{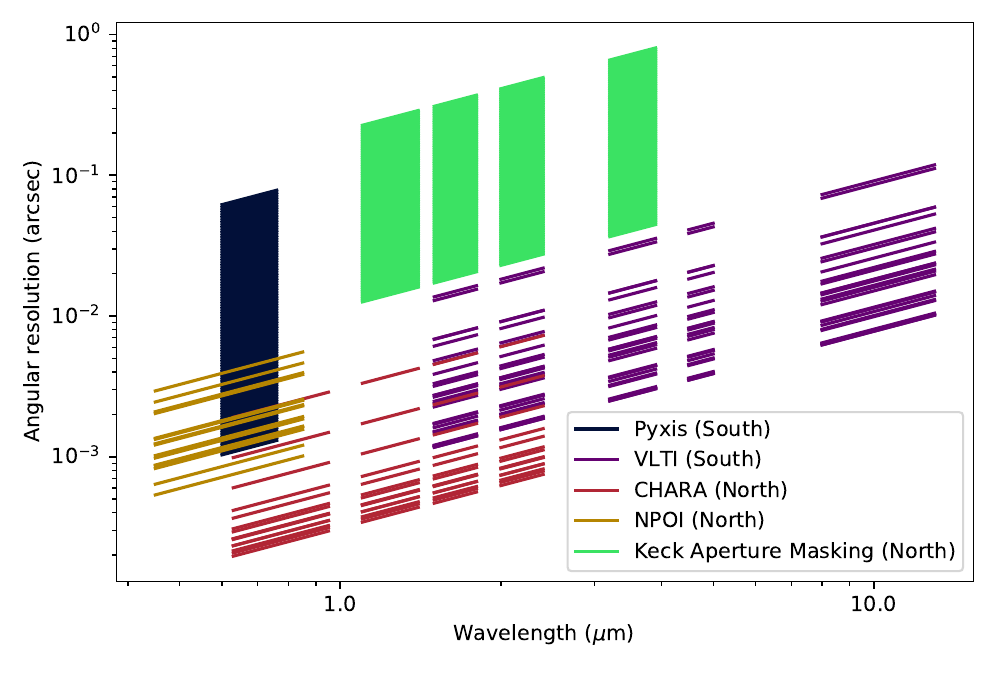}
    \caption{Comparison of the wavelength-angular resolution space of \textit{Pyxis} and current interferometric facitilites. To create this plot, the most popular and accessible baselines were chosen for each facility, noting that both NPOI and VLTI have a much wider range of baselines that are not normally available. The phase space of the Keck Aperture Masking experiment is also included. Angular resolution is calculated as $0.5\lambda/B$. Of particular note is that \textit{Pyxis} spans a unique portion of this phase space for the Southern Hemisphere.}
    \label{fig:interferometer_comparison}
\end{figure}

With \textit{Pyxis} working in the $R$ band, it is well placed to complement the existing suite of interferometers globally; particularly as it will be the only visible light interferometer in the Southern Hemisphere following the decommissioning of the SUSI interferometer \cite{Davis1999} in the mid-2010s. Figure \ref{fig:interferometer_comparison} shows the current angular resolution capabilities of the VLTI in Chile, and NPOI \cite{Armstrong1998} and the CHARA array in the USA as a function of wavelength, compared with \textit{Pyxis}. We also include the non-redundant baselines of the Keck Aperture Masking experiment \cite{Tuthill2000}, as a comparison for short baseline capabilities. As can be seen, \textit{Pyxis} will span a unique combination of wavelength and angular resolution within the southern hemisphere, and the ability for it to span a continuous range of baselines and position angles (and therefore uv plane data points) can allow it to probe optimal spatial frequencies for a given object's visibility curve.

One of the key areas that \textit{Pyxis} will provide insight into is the measurement of fundamental parameters of stars. In recent years, obtaining the stellar masses, ages and radii of stars with high precision is especially important due to the burgeoning fields of Galactic archaeology and exoplanet research, where the properties of the planet host star are critical in extracting the exoplanet parameters \cite{Clark2021,Rains2021,Tayar2022}. \textit{Pyxis} will build off of the success of PAVO in measuring stellar diameters \cite{White2013} utilising an even simpler architecture, single mode spatial filtering, and polarisation control. 

Precision stellar masses are another key parameter in the studies of stellar evolution, as they determine stellar age and Galactic evolution timescales. This parameter can be obtained through giant star asteroseismology, but suffers from a lack of calibration and benchmarks \cite{Epstein2014,Valentini2019}. Simple stellar diameter measurements have been shown to help calibrate these techniques \cite{Huber2012}, and so \textit{Pyxis} will be able to augment Gaia astrometry of the brightest astrometric binaries with precise interferometric separation measurements, thus providing a precise stellar mass with which we can calibrate other measurements.

Arguably the more exciting and unique science case of \textit{Pyxis}, however, is utilising its polarimetry functionality. First pioneered by Ireland (2005) \cite{Ireland2005}, multi-wavelength interferometric polarimetry has resulted in exciting observations in resolving spherically symmetric dust shells around very large stars that would be otherwise undetectable with a non-interferometric instrument \cite{Norris2012}. Such studies have probed the dust grain size distribution around giant stars, which in turn informs the processes of how dust is made and stellar mass loss. However, these studies have been rare due to the lack of available instrumentation to make these measurements, and questions remain regarding when dust scattering provides the dominant mechanism for giant star mass loss \cite{Hoefner2018}. 

\textit{Pyxis} requires a polarisation split in order to accurately calibrate the visibilities, and due to its much simpler geometry than existing long baseline interferometers, it is straightforward to split the polarisation for scientific measurements; \textit{Pyxis} should achieve an estimated calibrated differential polarisation fringe visibility of 2\% precision, comparable to that of previous measurements \cite{Norris2012}. Hence \textit{Pyxis} should be able to make simple time and wavelength-dependent interferometric polarimetry measurements around these bright giant stars to resolve some of these questions.

%\red{[I think it would be good to have a separate section or paragraph on the different technical/system elvel/operational requirements of Pyxis to achieve the above mentioned science goals. This includes, the metrology accuracy, angular positioning accuracy, fringe visibility, overall sensitivity , etc... I think we have captured in the confluence page: This could be a table as well...]}
%%%%%%%%%%%%%%%%%%%%%%%%%%%%%%%%%%%%%%%%%%%%%%%%%%%%%%%%%%%%%%%%%%%%%%%%%%%%%%%%%%%%%%%%%%%%%%%%

\section{Mechanical Design}
\label{sec:mechanical}
\subsection{Robotic Platforms}
\label{sec:robots}
The \textit{Pyxis} robotic platforms comprise a vibration-isolated upper platform, where lasers, telescopes, cameras, and fibre injection systems are mounted, and a lower platform housing the beam combiner, circuitry, and control computers. The upper platform payload on each robot is mounted on a stepper-motor-controlled goniometer to achieve precise elevation control at the level of a few arcseconds. These goniometers are coupled to the upper platform through a set of passive mechanical vibration isolators, intended to attenuate vibrations from roughness in the surface the robot traverses, as well as from the motors themselves. 

\begin{figure}
    \centering
    \includegraphics[width=0.8\columnwidth,angle=0]{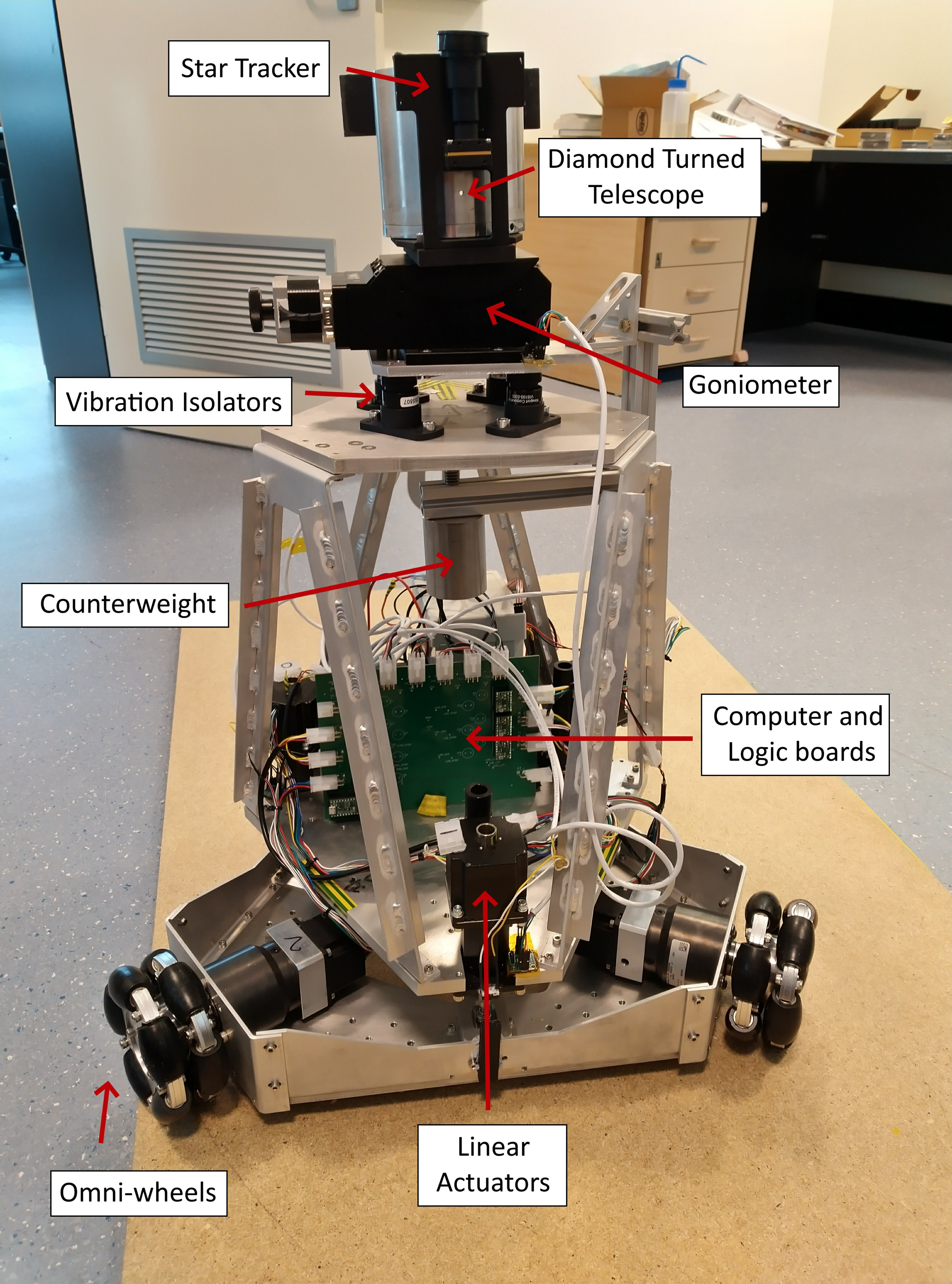}
    \vspace{0.5cm}
    \caption{A photograph of the deputy platform. The upper platform containing a goniometer, telescope and star tracker is shown above vibration isolators. On the chief platform, this upper platform also contains the fibre injection unit (see Section \ref{sec:BC}) and fibre optic gyroscope.}
    \label{fig:robot}
\end{figure}

The platforms are designed to allow control of all six degrees of freedom of the upper platform payload as a ground based simulation of satellite control.  The three degrees of freedom in the ``ground'' plane (two horizontal translations and rotation about the vertical axis) are controlled by stepper motors, coupled to precision planetary gearboxes and bi-directional omni-wheels in a ``Kiwi drive'' arrangement.  The three remaining degrees of freedom (vertical translation, tip and tilt) are controlled by a set of three linear actuators (also stepper-motor driven) with a three way rotationally symmetric, ball and V-groove kinematic interface.  The V-grooves are created by pairs of hardened dowel pins, creating a pair of point contacts with the truncated and threaded balls attached to the ends of the linear actuators. A photograph of one of the deputy platforms is shown in Figure \ref{fig:robot}.

\red{We note at this point that the robotic platforms do not provide a one-to-one reproduction of the type of formation flying that occurs in space, as they are co-located on the ground with only 5\,cm of vertical motion out of the plane. This means that the open-loop paths of the robots on the ground is different, but we are able to use the full 6-axis control to simulate different thruster actuators in space, using the same sensor suite that would be used for the closed loop control in space.} 

%The telescope is made of a 81~mm high cylinder hosting the 25~mm secondary mirror and attached to the cell of the primary mirror which has a 94~mm clear aperture.

Within the optical subsystems, fine tip/tilt control is achieved using piezo actuators and optical path difference is controlled with a piezo stick/slip stage. In order for the system to achieve stable fringes using these fine control subsystems, we determined the following mechanical requirements for the robotic platforms:

\begin{itemize}
    \item The RMS motion above 100~Hz for anything on the upper platform goniometer must be below 50~nm, so that fringe visibility remains high in 5\,ms exposures.
    \item The RMS velocity must not exceed 10~\textmu m/s at frequencies $<$100\,Hz, so that with a 5\,ms servo lag, fringe tracking at this 50\,nm level is possible, with residual fringe motion being dominated by astronomical seeing.
    \item The absolute positioning must be accurate to within 3~mm, in order to have a 1 part in 1,000 baseline knowledge for a 3\,m science baseline, and in order to keep the fringes within the range of the position actuator.
    \item The attitude of the platforms must be accurate to within 30", in order to inject the light into the field of view of the injection unit.
    \item The attitude velocity error cannot exceed 100"/s, so that with a 5\,ms servo lag, there is no more than a 0.5" angular error (required for single mode fibre injection).
\end{itemize}

\begin{figure}
    \centering
 %   \begin{subfigure}{0.7\textwidth}
 %   \centering
 %   \includegraphics[width=\columnwidth]{figures/Bode_linear.pdf}
 %   \caption{Linear scale}
 %   \label{fig:linear_sim_resonance}
 %   \end{subfigure}
 %   \begin{subfigure}{0.7\textwidth}
 %   \centering
    \includegraphics[width=0.8\columnwidth]{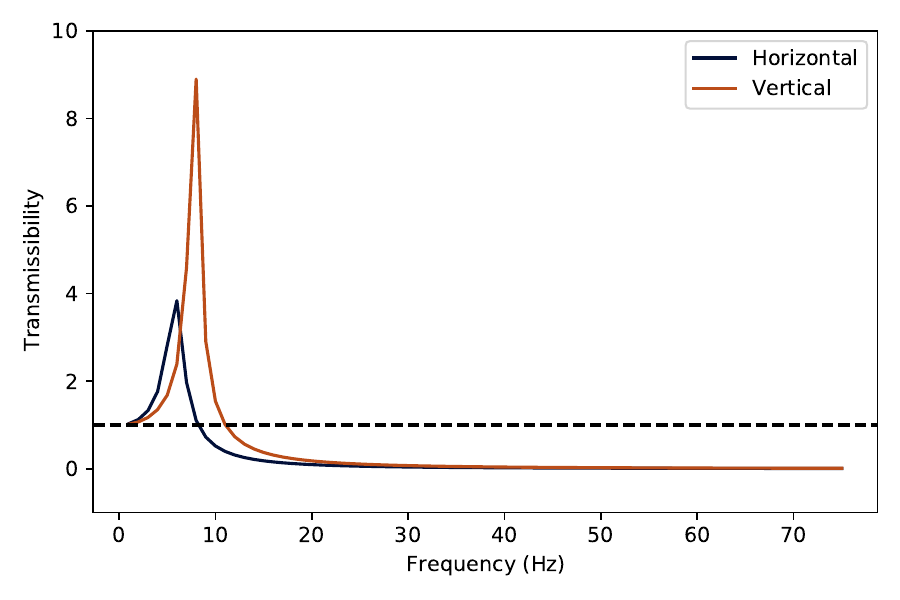}
 %   \caption{Logarithmic scale}
 %   \label{fig:log_sim_resonance}
 %   \end{subfigure}
 %   \vspace{0.3cm}
    \caption{Theoretical frequency/resonance response of the platforms for \red{the horizontal and vertical} axes of motion in terms of transmissibility, the ratio of \red{output amplitude to input} amplitude. Generated through fitting a damped oscillator to the parameters of the upper platform's springs. Black dashed line indicates a transmissibility of one.} %Inset in Figure 4.a shows a zoomed in view of frequencies between 0 and 20~Hz. Plot is scaled to allow a direct comparison with the measured response in Figure \ref{fig:resonance}.}
    \label{fig:sim_resonance}
\end{figure}

Since precise positioning is essential for the successful operation of the optics, the mechanical vibration isolators were characterised both through simulation and testing. The simulated transmissibility plot of their frequency response in the horizontal and vertical directions is displayed in Figure \ref{fig:sim_resonance}, produced by fitting a damped oscillator function to the listed parameters of the upper platform springs. The resonance peaks with over-unity gain in the 5-10 Hz range will cause fringe motions that are measurable and partly controllable by the fringe tracker operating at a 200~Hz loop speed. From these peaks, attenuation \red{improves over the $\sim$10-75~Hz range}, and it is expected that vibrations exceeding 75~Hz will not be of much concern. It is also clear that the resonance in the vertical direction has greater gain, as well as a higher frequency, and so is likely to be more of an issue than the horizontal resonance. Simulations were also carried out on a two-dimensional model of the system, and showed how horizontal motion would cause significant angular perturbation of the platform given the high centre of mass, leading to the installation of a 1kg counterweight suspended below the upper platform. 

\begin{figure}
    \centering
%    \begin{subfigure}{0.7\textwidth}
%    \centering
%    \includegraphics[width=\columnwidth]{figures/Transmissibility_linear.pdf}
%    \caption{Linear scale}
%    \label{fig:linear_resonance}
%    \end{subfigure}
%    \begin{subfigure}{0.7\textwidth}
%    \centering
    \includegraphics[width=0.8\columnwidth]{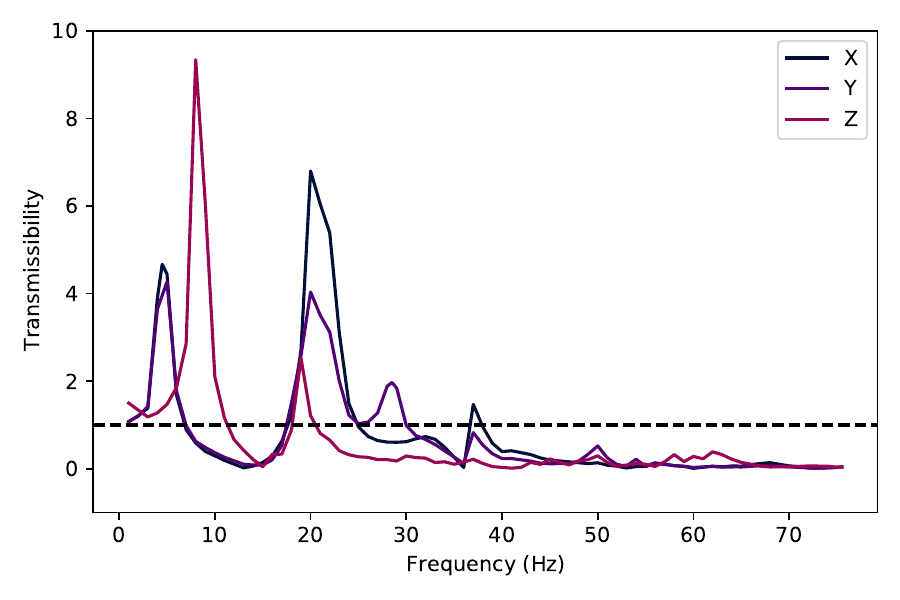}
%    \caption{Logarithmic scale}
%    \label{fig:log_resonance}
%    \end{subfigure}
%    \vspace{0.3cm}
    \caption{Measured frequency/resonance response of the platforms for various axes of motion in terms of transmissibility, the ratio of \red{output amplitude to input} amplitude. Black dashed line indicates a transmissibility of one.} %Inset in Figure 5.a shows a zoomed in view of frequencies between 0 and 20~Hz. }
    \label{fig:resonance}
\end{figure}

The system response between the lower and upper platform was physically tested, by applying a sinusoidal frequency sweep between 0 and 75~Hz using the motors in each axis (X, Y, Z), and recording the data from a set of three 3-axis accelerometers placed on the vibration-isolated platform. The accelerometer readings were bias-corrected and transformed into body-frame accelerations, before being Fourier transformed to find the peak amplitude for the given test frequency. The transmissibility -- the ratio of \red{upper platform (output) to lower platform (input)} amplitudes -- is plotted for all \red{three translational} axes of motion in Figure \ref{fig:resonance}. Between 1-15~Hz, these results largely agree with the simulation, showing peaks close to where they were expected for horizontal (X/Y) and vertical (Z) inputs, and significantly higher gain in the vertical direction. Unlike the theoretical plot, however, we do see some amplification of frequencies around 20-30~Hz in all three axes. These resonances are likely due to the coupling of the full system to the springs, and while the addition of a counterweight was able to suppress these resonances they were not able to be removed completely. Beyond \red{$\sim$30-40~Hz}, once again the springs attenuate to below unity gain.

% include discussion of simulated vibration resonances and experimental results from resonance tests

It is clear from both model and test results that the vibration-isolators alone are insufficient to attenuate vibration to the required levels over the full range of frequencies. However, given their good performance at high (\red{$>$40~Hz}) frequencies, it is expected that an active control system will be able to handle the lower frequency vibrations, discussed further in Section \ref{sec:control}.

\subsection{Diamond-turned Telescope}
\label{sec:telescope}
% Talking points: novelty of telescope design, how does it function, pretty pictures?
% Led by: SW, MI

\begin{figure}
    \centering
    \includegraphics[width=0.7\columnwidth]{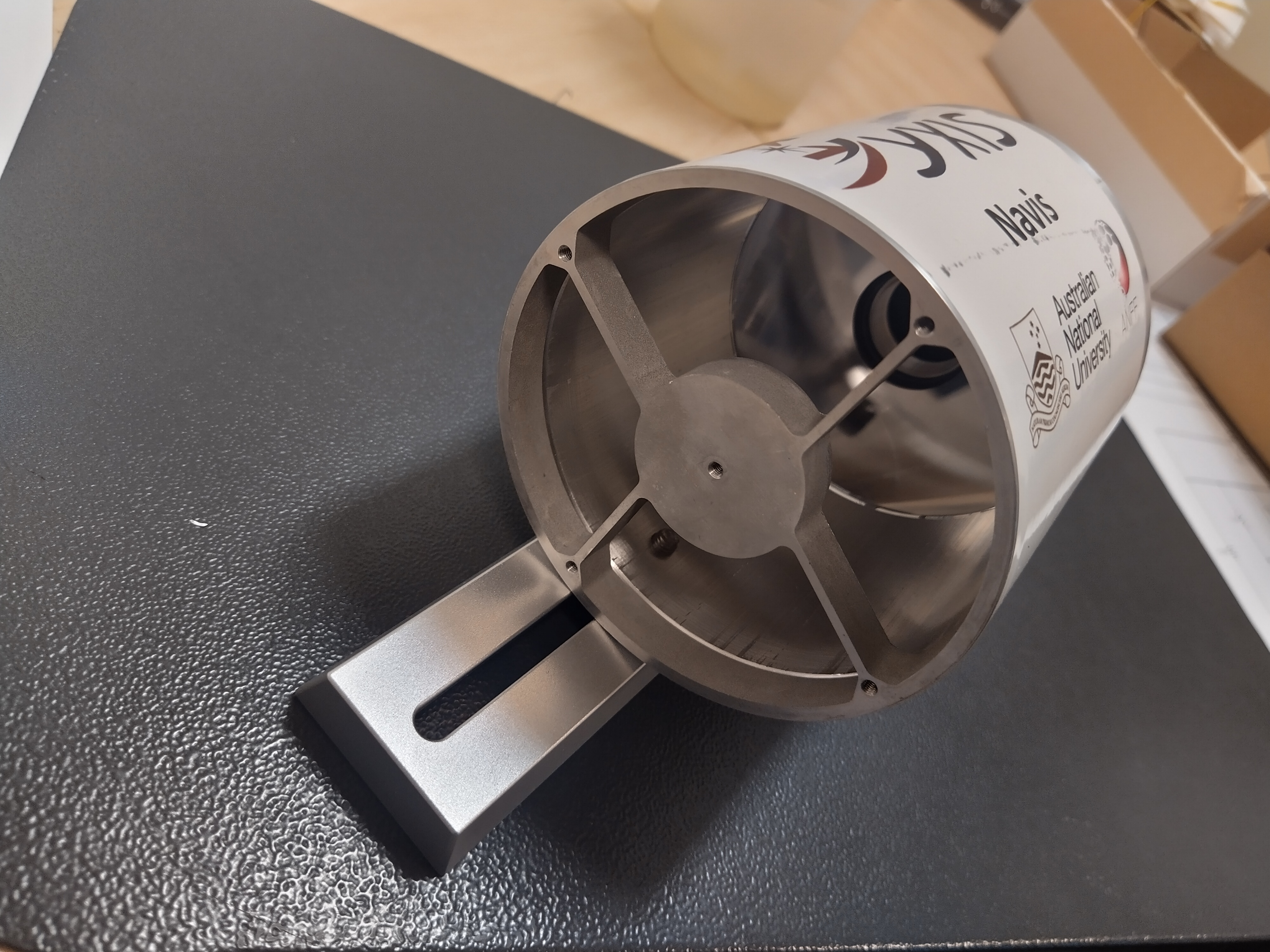}
    \vspace{0.5cm}
    \caption{A photograph of one of the diamond-turned aluminium telescopes.}
    \label{fig:telescope}
\end{figure}

The science telescopes are one of the subsystems that the project designed to be ``space ready''. The optics need to be significantly compressed to be compatible with the CubeSat format without need for refocusing, while sustaining the standard NASA \red{General Environmental Verification Standard (GEVS)} vibration qualification profile. Current demonstration units have passed this requirement. In addition, it requires a wide enough field of view so that it could correct angular errors in the deputy position without moving parts. Our solution was to design and prototype the telescopes from diamond-turned aluminium. A photograph of one of the telescopes is shown in Figure \ref{fig:telescope}.

%\red{[This shock level seems to be very high, is it pyroshock? What kind of vibration we are talking here, random vibration? The standard NASA GEVS qualification level random vibration value is ~ 14.1g RMS. Whatever HEO is doing]}

Each deputy telescope has a 5:1 magnification conjugated at infinity, in a Cassegrain design with two paraboloids. The primary has a 94~mm clear aperture diameter with a 200~mm radius of curvature, while the secondary is 25~mm across with a 40~mm radius of curvature. This size allows the telescope to fit in one end of a 3U CubeSat, and can be seen attached to the upper platform in Figure \ref{fig:robot}. A 45 degree flat tertiary mirror reflects the beam at 90 degrees to the optical axis.  The telescopes were manufactured at Optofab-ANU, part of the Australian National Fabrication Facility, using a diamond lathe turning RSA-6061 aluminium. The complete telescope structure, including mirrors, is formed from aluminium, so that the telescope is naturally resistant to optical aberrations caused by thermal expansion. 

The telescope manufacturing and assembling process went through several prototypes, \red{where our overall requirement of $<$100\,nm RMS wavefront error excluding focus consists of a $<$60\,nm RMS wavefront error from each of the primary and secondary, and a $<$50\,nm RMS wavefront error from telescope alignment and stress, and contributions from the flat secondary mirror. The focus requirement for the telescope is derived from a negligible loss in coupling over the full baseline range, resulting in a very tight $<$50\,nm RMS requirement for the focus term.} We have minimised stress on the primary mirror caused by the assembly of the two parts of the telescope: we moved away from an original shrink fit design and have developed an assembly that minimises distortions by using a 50 micron thick gap around the edge of the primary mirror filled with a thermal-expansion matched vacuum compatible adhesive. Coma and defocus are actively set to have negligible amplitudes interferometrically during adhesive curing, by fine adjustment of primary mirror tilt and piston. While a residual amount of spherical aberration remains, along with some minor astigmatism caused by the gluing and mounting of the tertiary mirror, the final complete telescopes exhibit a $\sim$70\% Strehl ratio, consistent with the $<$100\,nm RMS wavefront error. \red{This reduces the coupling into the fibres, amounting to a 0.3 magnitude loss in sensitivity.}

%%%%%%%%%%%%%%%%%%%%%%%%%%%%%%%%%%%%%%%%%%%%%%%%%%%%%%%%%%%%%%%%%%%%%%%%%%%%%%%%%%%%%
\section{Metrology System}\label{sec:metrology}

One of the primary goals of \textit{Pyxis} is to demonstrate a metrology system capable of sustaining satellite formation flight at an adequate level to make precise interferometric measurements. \textit{Pyxis}'s distance metrology approach is divided into two broad parts: a coarse metrology system that measures the \red{platform positions with respect to the chief in 3 dimensions}, and a fine metrology system that complements the coarse metrology using interferometry and fringe patterns to achieve the necessary sub-wavelength precision. The fine metrology system concept was already described in Lagadec et. al (2020)\cite{Lagadec2020}. This system is not yet fully commissioned, in particular with a decision point remaining as to whether temperature stabilised Fabry-Perot laser diodes are enough, or if three single frequency laser diodes are required. However, the combination of coarse metrology described below, combined with a fringe search (see Section~\ref{sec:fringe_tracking_control}) using starlight will be enough for initial on-sky fringes.

\subsection{Coarse Metrology}\label{sec:coarse_metrology}

% Talking points: LED/TOF Metrology subsystem (Patrick/Joice's work)
% Led by: JM, PM

The coarse metrology itself has two parts: measuring the \red{angular position of two LEDs, including the} parallax between the two LEDs, and a time-of-flight (TOF) laser system. The \red{first part} consists of two LEDs mounted \red{vertically}, 90~mm apart on the sides of each of the deputy platforms, and a camera mounted on the chief robot.
We measure the angular separation of these LEDs to 0.05 pixels or 0.7" using a 7.6~mm focal length F/2 lens attached to a FLIR Firefly FFY-U3-16S2M-S camera. This in turn will enable a \red{distance} measurement of \red{$\sim$1\,cm for a chief-deputy separation of 10~m}. \red{We note here that the distance measurement is differentiated from any rotation of the platforms through the use of a star tracker, which constrains the robot's orientation to within ten arcseconds (see Section \ref{sec:startracker}). Additionally, this 0.7" LED angular position, when combined with the accurate on-axis position of the chief platform, results in a deputy position knowledge along the star vector to within $\sim$0.1\,mm. The deputy position out of the plane of Figure \ref{fig:pyxis_schematic} (orthogonal to baseline and star vectors) is dominated by the maximum 30" angular uncertainty of the chief about the star vector (solved for using other stars in the field), resulting in a maximum 5\,mm out of plane position uncertainty for the 60\,m baseline.}

However, \red{to achieve \textit{Pyxis'} nominal baselines of $\sim$60~m, and for the purposes of designing a future space-based mission with a baseline of $\sim$300~m, this camera-based metrology system is almost certainly insufficient, even with a potential extended LED spacing of 300~mm.} This is due to the precision required for angular separation measurements, which is \red{inversely} proportional to the square of the spacecraft separation. Hence, while we implement this system for \textit{Pyxis}, the TOF system is more critical for demonstrating a complete space compatible metrology system. The TOF system is also required to obtain a sub-centimetre position estimate, in order to be within the fine metrology system's capture range.

The TOF distance ranging system utilises the time it takes for light to travel from the chief platform to a retroreflector mounted on a deputy platform and back. Pulses of light are generated by a laser diode and the distance the light travels determines the delay before the reflected light returns. This is equivalent to a phase shift in the pulses of the returning light relative to the transmitted pulses. The distance to the reflecting object, $d$, is half the total distance the light has travelled and is related to the delay in the returning light pulse $t_{\rm d}$ by the speed of light, $c$. Hence this can be expressed as $d= t_{\rm d} \cdot \frac{c}{2}$.

Implementing a TOF system with analogue electronics requires two capacitors that act as timing elements to measure the delay and a clock source to generate pulses of light. The same clock pulses are used to switch between the two timing elements so that the first capacitor is charged while the pulse is high and the second capacitor charges while the pulse is low. When the capacitor is connected, it is charged by a photo-diode which allows a current to flow when it receives the returning pulses of light, charging the connected capacitor. An FPGA with an integrated microcontroller and some additional circuitry including the laser, photodiodes, timing capacitors, fast analogue switches and an analogue-to-digital converter (ADC) is used to implement the system. 

A high frequency, square wave signal is produced by clock conditioning circuity within the FPGA fabric. Two signals with the same frequency are produced, one to drive the laser modulator and the other to drive the switching between timing elements. The timing is implemented using a pair of capacitors that are charged by the photocurrent from a photodiode detector and a high-speed analogue 2:1 multiplexer (MUX) is used to direct the current between the two charge-integrating capacitors. After an integrating cycle, the capacitors are sequentially sampled by an ADC, which has a built-in MUX, and the capacitors are reset by a command from the microcontroller.

In the full system, there are four pairs of capacitors and accompanying photodiodes, so that the system can measure two distances to each of the deputy platforms. The light from a single modulated laser diode is optically split for the four light paths. All four channels are charged in parallel and then sampled sequentially. The laser modulator is controlled by the clock signal and provides the required DC bias and modulation current to drive the laser diode. The laser diodes used in this system are shared with the fine metrology system (described next in Section \ref{sec:fine_metrology}), with the laser diodes and TOF photodiodes shown in the metrology schematic in Figure \ref{fig:metrology_schematic}.

\subsection{Fine Interferometric Metrology}\label{sec:fine_metrology}

% Talking points: fine metrology subsystem developments (Tiph and Adam's work, refer to Tiph's 2020 SPIE paper)
% Led by: AR, MI?

To achieve sub-wavelength precision metrology, we require the use of interferometric fringe measurement. \red{The measurement is made to a retro-reflector on the deputy spacecraft, which is in turn locked to the pivot point of the deputy through the deputy's star tracker. In principle, a predicted fringe motion error up to 5\,$\mu$m could be caused by motion of this retro-reflector vertex with respect to the deputy pupil location. We do not intend to take this into account in the first instance, and will consider the retro-reflector vertex to have a fixed offset to each deputy pupil.} However, interferometric metrology itself relies on the coherence of its beams to the sub-wavelength level. Such required precision poses a challenge for a multiple platform system like \textit{Pyxis} for which fringe scanning - necessary to unambiguously determine the phase and optical path length difference - is complex and time consuming. Fortunately, it is possible to employ the \red{technique} of \textit{Multi-Wavelength Interferometry}\cite{Polhemus1973} to broaden the wavelength range where the phase can be unambiguously measured; using several wavelengths can broaden the range by three orders of magnitude compared to the narrow $\pm\frac{\lambda}{2}$ possible with a single wavelength. The details of this technique as applied to \textit{Pyxis} are described in Lagadec et. al (2020)\cite{Lagadec2020}, but in brief a `synthetic' wavelength is created between each combination of wavelengths, given by:
\begin{equation}
    \Lambda = \frac{\lambda_1\lambda_2}{(\lambda_2-\lambda_1)},
\end{equation}
for two wavelengths $\lambda_1$ and $\lambda_2$. Hence, as long as the coarse metrology can provide an estimate of the distance to within the range of the longest synthetic wavelength, we can provide a precise measurement using this system. 

\begin{figure}
    \centering
    \includegraphics[width=\columnwidth]{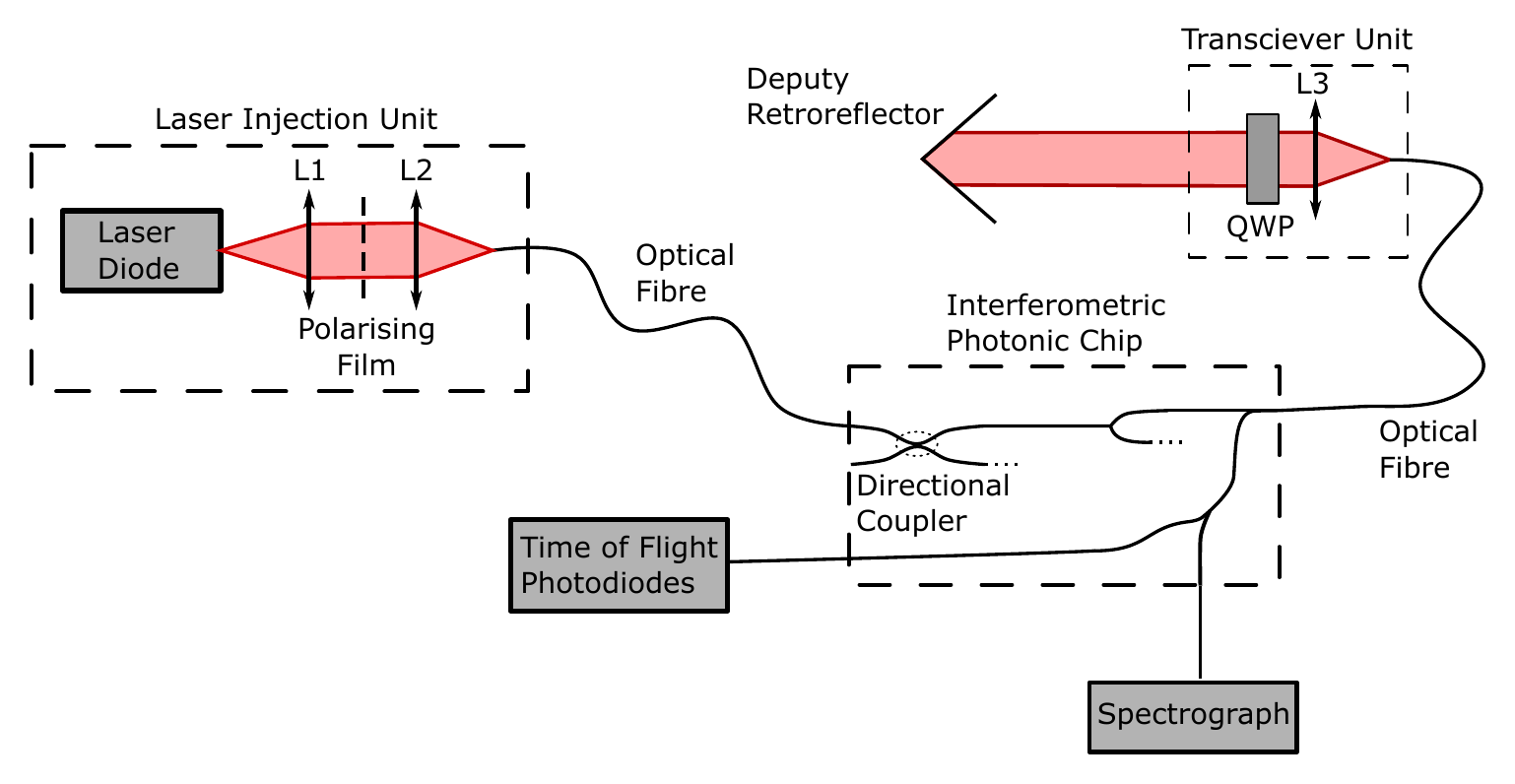}
    \vspace{0.3cm}
    \caption{Schematic of the metrology system as described in Section~\ref{sec:fine_metrology} and adapted from Lagadec et. al (2020) \cite{Lagadec2020}. Note that this schematic only shows one out of the two laser diodes, and one of the four beam paths through the chip towards the deputy retro-reflectors. The laser diodes are also used for the time-of-flight coarse metrology (see Section \ref{sec:coarse_metrology}), and the photodiodes displayed are specifically included for this system.}
    \label{fig:metrology_schematic}
\end{figure}

\begin{figure}
    \centering
    \includegraphics[width=0.9\columnwidth]{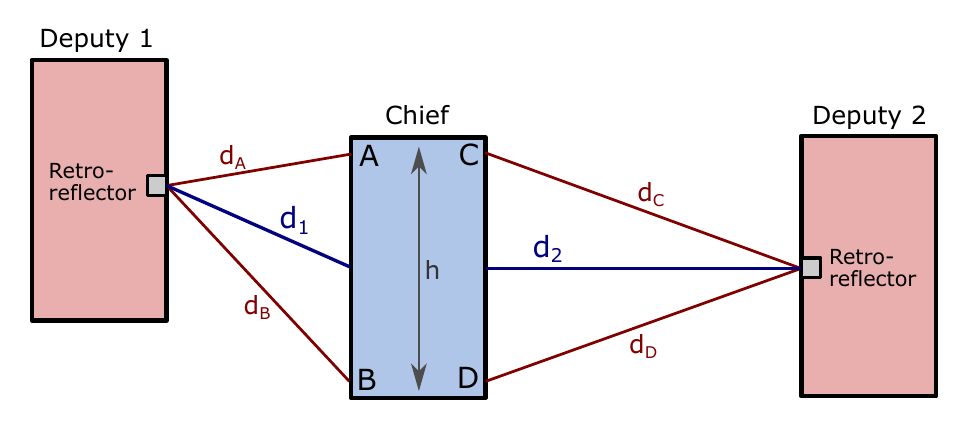}
    \vspace{0.3cm}
    \caption{Metrology architecture adapted from Lagadec et. al (2020)\cite{Lagadec2020}. Ultimately, the metrology system is designed to measure the differential distance between the deputies ($d_1 - d_2$) to sub-wavelength precision. The light from the metrology is emitted from transceiver units located at positions A, B, C and D. The light is then retro-reflected from the deputies back into the metrology system in the chief, providing four distance measurements \red{$d_A$, $d_B$, $d_C$ and $d_D$}.}
    \label{fig:metrology_measurement}
\end{figure}

To implement this in \textit{Pyxis}, we use a similar setup to Lagadec et. al (2020)\cite{Lagadec2020}, which is shown in Figures  \ref{fig:metrology_schematic} and \ref{fig:metrology_measurement}. First describing Figure \ref{fig:metrology_schematic}, the light from each diode is collimated by a 4.5~mm focal length lens (L1) and injected into an optical fibre by a 6.25~mm focal length lens (L2). The injection unit involves an in-house constructed diode-to-fibre injection system that couples into an APC fibre connector of a polarisation maintaining fibre via a polarising element. This results with a high ($\sim$30\,dB) polarisation extinction ratio into one of the orthogonal fibre modes. Note the schematic only shows one of the two laser injection units.

The fibres are fed through a V-groove into a photonic chip, where the light is mixed by a directional x-coupler. This chip is currently being manufactured through Australian National Fabrication Facility OptoFab node through direct write ultrafast laser inscription (ULI) \cite{Osellame2012,Gross2015}, and has the advantage of being able to combine elements of the coarse and fine metrology systems in a compact form factor ideal for a future space mission.

The light is split into four separate beams (only one beam path is shown in Figure \ref{fig:metrology_schematic}) and output into a series of optical fibres. These are then fed into four transceiver units, which consist of a 25~mm focal length collimating lens (L3) and a quarter-wave plate. The collimated light is sent out of the chief platform at positions A, B, C and D shown in Figure \ref{fig:metrology_measurement}, reflected off of a retro-reflector mounted on each deputy platform (element 10 in Figure \ref{fig:pyxis_schematic}), and back into the transceiver unit. \red{The four laser measurements are used to measure the distance in two dimensions, with the third dimension (orthogonal to both the baseline and the star vector) not requiring sub-micron precision.} The quarter-wave plate ensures that the outgoing and incoming polarisations are orthogonal and will not interfere with each other.

A test configuration using two Fabry-Perot laser diodes as injection units has been constructed. Continued development of this system, including thermal stabilisation of the Fabry-Perot diodes and investigation of single-frequency diodes is ongoing.

%%%%%%%%%%%%%%%%%%%%%%%%%%%%%%%%%%%%%%%%%%%%%%%%%%%%%%%%%%%%%%%%%%%%%%%%
\section{Injection and Beam Combination}
\label{sec:BC}
\subsection{Fibre Injection}
\label{sec:FI}
% Interferometric Beam Combiner
With \textit{Pyxis} being a relatively simple, single baseline interferometer, we aimed to produce a beam combiner that was as simple as possible to maximise throughput. We also aimed to have limited moving parts, to maximise the amount of coherent integration. This will also serve well in translating this beam combiner design to a future space-based mission. In this vein, we chose to base the beam combiner around an integrated optics (IO) photonic chip modelled after the success of IO chips in the GRAVITY \cite{Abuter2017} and GLINT \cite{Norris2020} instruments.

\begin{figure}
    \centering
    \includegraphics[width=\columnwidth]{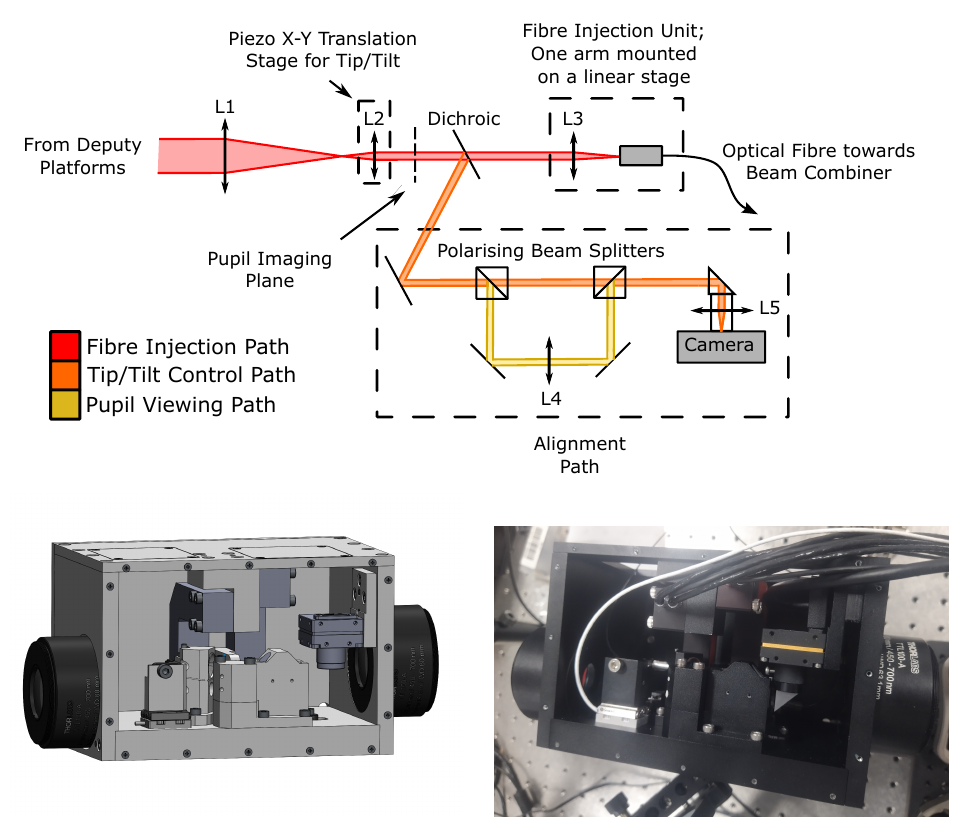}
    \vspace{0.5cm}
    \caption{\red{Top: Schematic of one of the two optical paths for the fibre injection system, as explained in Section \ref{sec:BC}. The various coloured beams denote different light paths of the subsystem: light redwards of 595~nm is injected towards the beam combiner (red), whereas the shorter wavelengths are sent into an alignment path, where one polarisation is used for tip/tilt correction (orange) of both beams, and the other polarisation images both pupils pupil (yellow). The alignment paths are imaged onto a camera within the assembly. Note that only one of the two fibre injection units are mounted on a linear stage for fringe tracking. 
    Bottom Left: CAD model of the fibre injection and alignment system. 
    Bottom Right: Photograph of the assembled fibre injection and alignment system.}}
    \label{fig:fibre_injection}
\end{figure}

The $\sim$18~mm diameter beams from each of the two deputy collector platforms are reflected towards the central platform and enter into the fibre injection system; shown in Figure \ref{fig:fibre_injection} for one of these light paths. This system is also designed to be CubeSat compatible, measuring 200 by 100~mm. The beam is transmitted through a $f=100$~mm focal length infinity corrected tube lens assembly (L1), then a piezo-controlled translating $f=6$~mm, 4~mm diameter collimating lens (L2), which acts as the tip/tilt actuator through X-Y translation orthogonal to the optical axis. We used Piezosystem Jena PXY 200 D12 stages, which when accounting for PWM filtering and voltage conversions, provides us with about 150~\textmu m of stroke.

A 595~nm cutoff dichroic is used to split the shorter wavelengths used for alignment from the longer science wavelengths, which are transmitted towards the fibre injection unit. The $<$595\,nm wavelengths are polarisation-split and recombined so that a single FLIR Firefly FFY-U3-16S2M-S camera can coarsely image both input pupils for pupil alignment with one polarisation while simultaneously performing fast ($\sim$200\,Hz sampling rate) tip/tilt with the other polarisation. The control system is explained in Section \ref{sec:tip_tilt_control}.

The pupil viewing path is designed such that the light passes through a $f=48$~mm lens (L4), which is focused at a pupil imaging plane located before the dichroic (seen in Figure \ref{fig:fibre_injection}). The camera images both paths through a $f=12$~mm imaging lens (L5) focused at infinity with a right-angled prism glued in front to ensure the camera fits within the tight footprint. It is also worth mentioning that while the footprint is tight, space inside the unit has been allocated to allow the insertion of a star tracking camera. This will be utilised for the CubeSat version of \textit{Pyxis}. 

\subsection{Science Beam Combiner}
The longer wavelengths, used for science, are injected into polarisation maintaining 630-HP fibres using two $f=4.5$~mm, 3~mm diameter achromat lenses (L3). One of the arms is mounted onto a small SmarAct SLC-1720-L translation stage, with 20~nm steps and 8~mm of stroke, for fine path control and fringe tracking. The fibres transport the light from the upper platform towards the bottom platform, where the beam combiner is located. The fibres are then fed into a V-groove, which is attached to the photonic chip. 

The chip features a ``tricoupler'' waveguide scheme, where three inputs are fed towards each other in an equilateral triangle formation, and then fed back out in three outputs. As described in Hansen et. al. (2022) \cite{Hansen2022}, the tricoupler results in the output beams having a phase shift of $\frac{2\pi}{3}$, and allows for the full recovery of the complex coherence in a single frame without modulation (see Sections 2.3 and 2.4 in that paper).  The three outputs that this chip provides is the minimum possible while retaining the above qualities of full information without modulation, and thus maximises throughput. The photonic chip was measured to have a mean throughput of 85 $\pm$ 7\%, and a coupling ratio between 33:33:33\% and 51:31:17\% over the science bandpass of approximately 620~nm to 760~nm. Because we only have two inputs from the telescopes, the central input of the chip was not used. More details regarding this photonic chip, including experimentally retrieved visibilities and group delay, and a schematic of the chip and attached V-groove, can be found in Hansen et. al. (2022) \cite{Hansen2022}. This paper also details the reduction algorithms using a pixel to visibility matrix ($\vb{P2VM}$), and group delay extraction for fringe tracking based on Fourier transform numerical integration. More details on the fringe tracking control loop can be found in Section \ref{sec:fringe_tracking_control}.

\begin{figure}
    \centering
    \includegraphics[width=0.6\columnwidth]{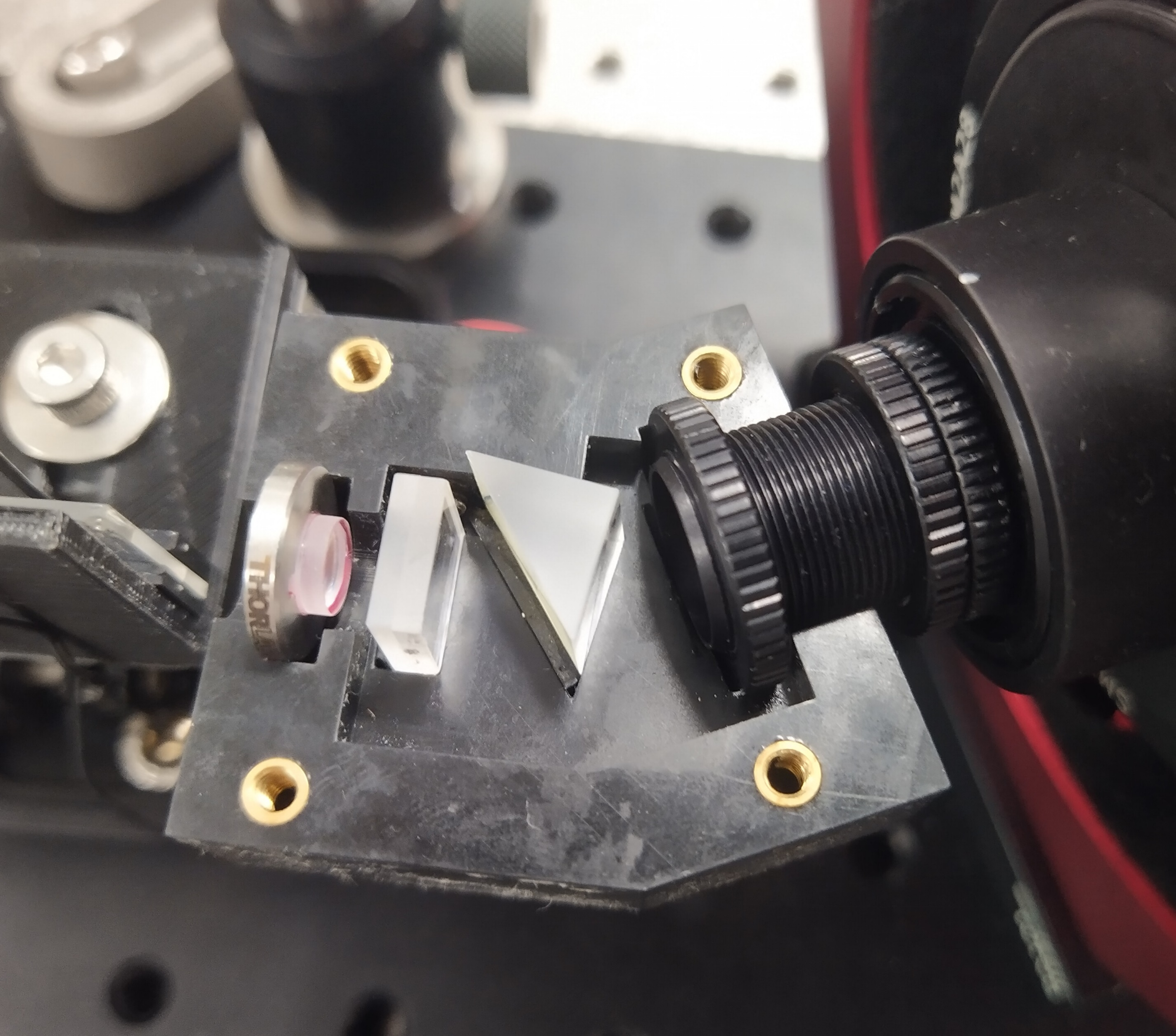}
    \vspace{0.5cm}
    \caption{3D printed resin mount for the spectrograph. From left to right, the optical components are: a $f=12.5$~mm collimating lens, a Wollaston prism, a 45$^{\circ}$ BK7 dispersion prism and a $f=15$~mm imaging lens. }
    \label{fig:spectrograph}
\end{figure}

From the beam combiner chip, the three outputs are fed through a $f=12.5$~mm, 6.25~mm diameter collimating lens into a custom spectrograph. This spectrograph includes a Wollaston prism for splitting linear polarisations (and hence allowing for both polarisation calibration and polarimetry on astrophysical sources) and a 45$^{\circ}$ BK7 dispersive prism. This results in the spectrograph having a spectral resolution of $R\sim 50$, which was chosen as a balance between throughput (i.e reducing the number of channels on the detector) and scientific usefulness. The dispersed light is then fed through an $f=15$~mm imaging lens onto the scientific camera. Due to the custom optical components and small space limitations, a resin printed optical mount was designed (shown in Figure \ref{fig:spectrograph}) to hold the spectrograph lenses and prisms. The mount contains a variety of screw holes and sprung inserts to ensure a kinematic mount of all components.

\begin{figure}
    \centering
    \includegraphics[width=0.8\columnwidth]{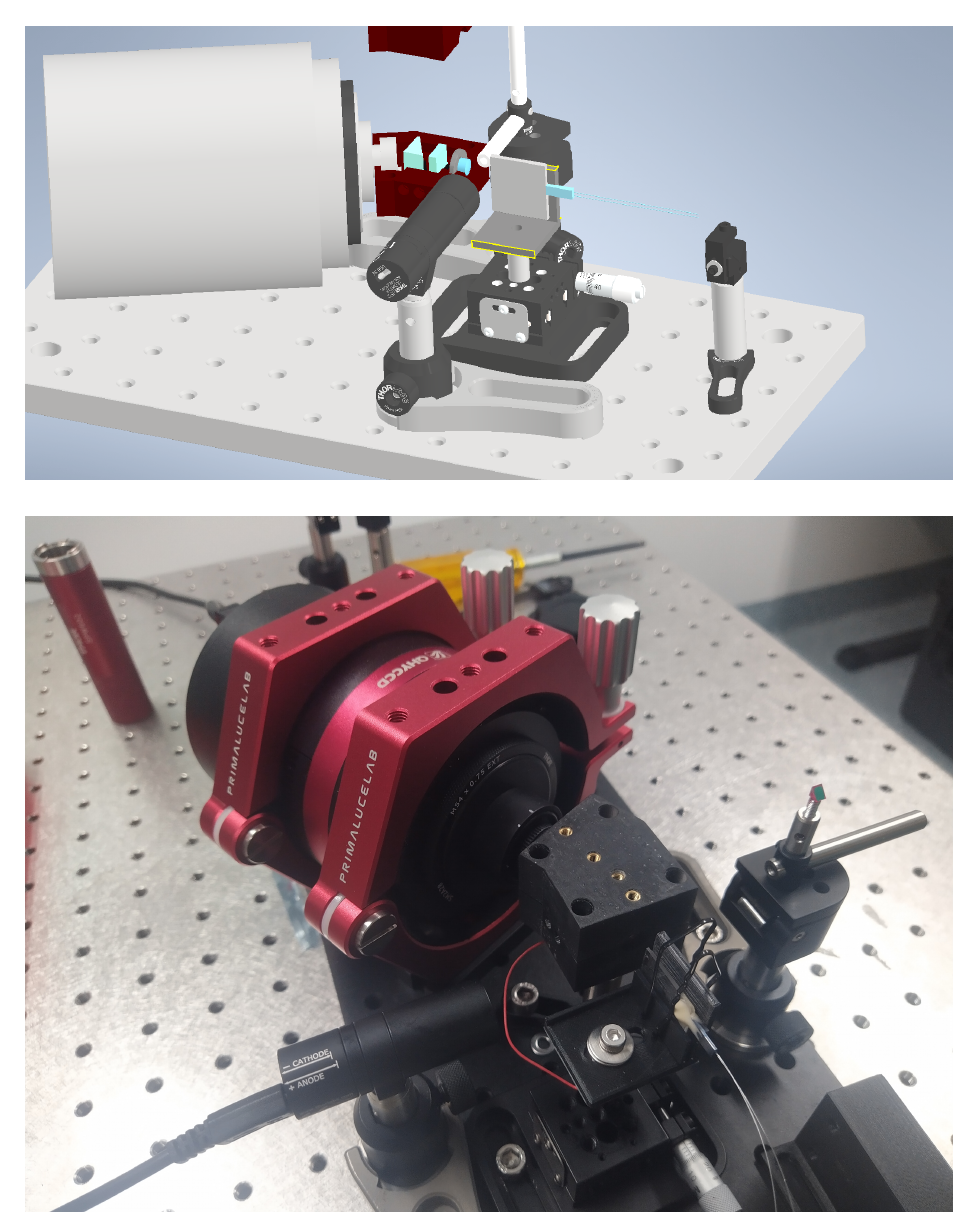}
    \caption{\red{CAD Design (top) and assembly (bottom)} of the science beam combiner, including the photonic chip, spectrograph, camera and back illumination optics.}
    \label{fig:beam_combiner}
\end{figure}

The full beam combiner design can be seen in Figure \ref{fig:beam_combiner}. The photonic chip is mounted to a manual translation stage for focus adjustment; due to the limited thermal expansion expected in the system, we do not anticipate needing to constantly adjust this axis of motion. The chip is also mounted atop a piezo stack to allow for small translations in the vertical direction; ensuring that the centre of the output beam lines up with detector pixels. 

The scientific detector chosen was a QHY2020 camera with the GPixel Gsense2020 BSI sensor. This was chosen for its low readout noise capabilities and high frame rate. The camera is mounted with a set of telescope tube rings and is connected to the spectrograph with a series of adapting rings. The whole system fits on a 150x300~mm breadboard, and in principle the size footprint could be made smaller through choosing a much smaller camera. The beam combiner also features a small back-illumination setup, consisting of a 590~nm LED that is reflected off of a miniature 45$^{\circ}$ mirror that is able to be flipped in and out of the optical path onto the output waveguides of the photonic chip. This will assist in fibre injection alignment, and show which pixel on the pupil tracking camera corresponds to the science fibres.

\subsection{Pupil Alignment Procedure}
\label{sec:system_alignment}

Before on-sky observations can be made, a variety of offsets and calibrations are required to ensure that the system can be easily aligned. Notable among these is a method to ensure an initial pupil alignment to within the capture range of the pupil camera. This comes from the coarse metrology system, with which we can identify a pixel offset from the two metrology LEDs corresponding to the exit aperture. \red{The alignment described below is solely for position corrections; any angular corrections are managed by the star trackers and tip/tilt sensors.}

\begin{figure}
    \centering
    \includegraphics[width=0.9\columnwidth]{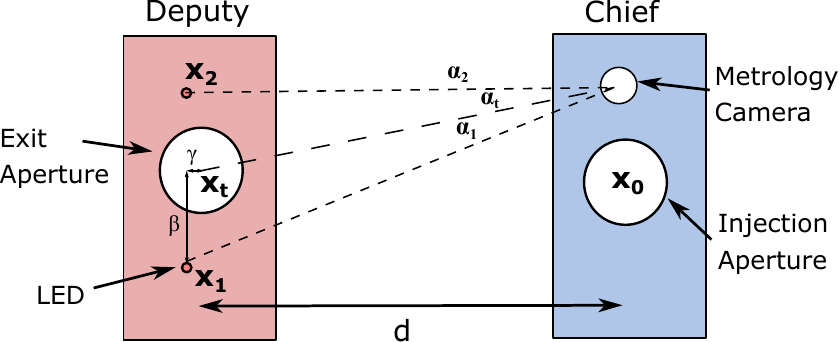}
    \vspace{0.5cm}
    \caption{\red{Schematic of the pupil alignment procedure. The location of exit and injection apertures are calibrated with respect to the two LED positions, which can be measured during operation via the metrology camera on the chief platform.}}
    \label{fig:met_align}
\end{figure}

Let us consider the two LEDs with coordinates $\vb{x_1}$ and $\vb{x_2}$ that are measured from the mechanical design, as well as the target exit aperture coordinate $\vb{x_t}$. These coordinates are a projection of the absolute coordinates onto the plane perpendicular to the baseline vector. We can then define a relationship between these three positions with respect to two parameters, $\beta$ and $\gamma$:
\begin{align}
    \vb{x_t} = \vb{x_1} + \beta(\vb{x_2} - \vb{x_1}) + \gamma \vb{R}_{90}(\vb{x_2} - \vb{x_1}),
\end{align}
where $\vb{R}_{90}$ is a 90 degree rotation matrix. \red{A schematic of these parameters and positions can be found in Figure \ref{fig:met_align}.}

On the coarse metrology camera, we then measure the two dimensional angles $\vb*{\alpha_i}$ corresponding to the two LEDs and can thus measure the angle of the exit aperture:
\begin{align}
    \vb*{\alpha_t} = (\beta + \gamma\vb{R}_{90})\vb*{\alpha_2} + (1-\beta -\gamma\vb{R}_{90})\vb*{\alpha_1}.
\end{align}
This angle can also be calculated when considering the frame of reference of the chief:
\begin{align}
    \vb*{\alpha_t} = \frac{\vb{x_0} + \vb*{\alpha_c}d}{d},
\end{align}
where $\vb{x_0}$ is the \red{injection} aperture coordinate also taking into account the offset from the chief metrology camera and $d$ is the distance between the two platforms. The angle $\vb*{\alpha_c}$ is \red{a calibrated correction for any angular deviation from the optical axis of the injection lens. This is calibrated in the lab using the back-illuminated beam from the science camera, with the tip/tilt piezos at the centre of their range.} The reference pixel is identified ahead of time through back illumination and set of retro-reflectors, and is the position on the tip/tilt sensing camera where the light is injected into the fibre. 

Hence, after calibrating the system, we simply need to adjust the deputy such that the LED angles $\vb*{\alpha_1}$ and $\vb*{\alpha_2}$ satisfy:
\begin{align}
    (\beta + \gamma\vb{R}_{90})\vb*{\alpha_2} + (1-\beta -\gamma\vb{R}_{90})\vb*{\alpha_1} = \frac{\vb{x_0} + \vb*{\alpha_c}d}{d},
\end{align}
while simultaneously measuring the distance $d$ through the approximation:
\begin{align}
    d \approx \frac{|\vb{x_1}-\vb{x_2}|}{|\vb*{\alpha_1} - \vb*{\alpha_2}|}.
\end{align}
Using the small angle approximation here is adequate for all platform separations exceeding 0.5m.

We mention here that while this calibration and alignment procedure will work for \textit{Pyxis}, it is insufficient for the space version; in a space environment, we are not able to move the deputies in all directions. Instead, we change the angle $\vb*{\alpha_c}$ by changing the central target positions of the tip/tilt piezo stages.

%%%%%%%%%%%%%%%%%%%%%%%%%%%%%%%%%%%%%%%%%%%%%%%%%%%%%%%%%%%%%%%%%%%%%%%%
\section{Control System}
\label{sec:control}
\subsection{Sensors and Architecture}

A key part of achieving the stability and positioning accuracy required for the interferometry is the navigation and control system. This system is comprised of a variety of sensors, actuators, and processing computers spread across the three robotic platforms and interfaces with all \textit{Pyxis} systems. Here we describe an overview of the physical elements that contribute to the control of \textit{Pyxis}, followed by the software architecture and the control system design. 

It is clear from Figure \ref{fig:pyxis_schematic} that many of the \textit{Pyxis} subsystems are devoted to metrology and navigation at varying levels of granularity in order to achieve sub-wavelength precision, with coarse sensors providing sufficient accuracy for unambiguous operation of higher resolution sensors. \red{However, we also note that the mechanical stability requirements in Section~2.1 represent the requirements for obtaining starlight fringes, which is possible without this nested measurement system.} Absolute attitude measurement is achieved using star tracker cameras on each deputy, described further in Section \ref{sec:startracker}. Attitude is also tracked via inertial measurements, using six 3-axis accelerometers on each robot, and a fibre laser gyroscope (FLG; VG035LND from Fizoptica) on the chief robot. The coarse metrology camera also captures the position of the deputy satellites in 3 dimensions with respect to the chief body frame. The central frame of reference, however, is defined through a high precision star tracker on the chief robot. This star tracker consists of a finite-conjugate version of the diamond-turned telescopes described earlier in Section \ref{sec:mechanical} combined with a FLIR Blackfly camera containing an IMX178 sensor. This allows us to obtain a $\pm 40$~arcminutes field of view with 1.5" per pixel, ample enough to sample the FWHM of a guide star under the seeing conditions of Mt Stromlo without pixel phase errors. This, together with the attitude measurements supplied from the sensors listed above, provides \textit{Pyxis} with a reference frame independent from the Earth and co-moving with the platforms themselves. 

\begin{figure}
    \centering
    \includegraphics[width=0.7\columnwidth]{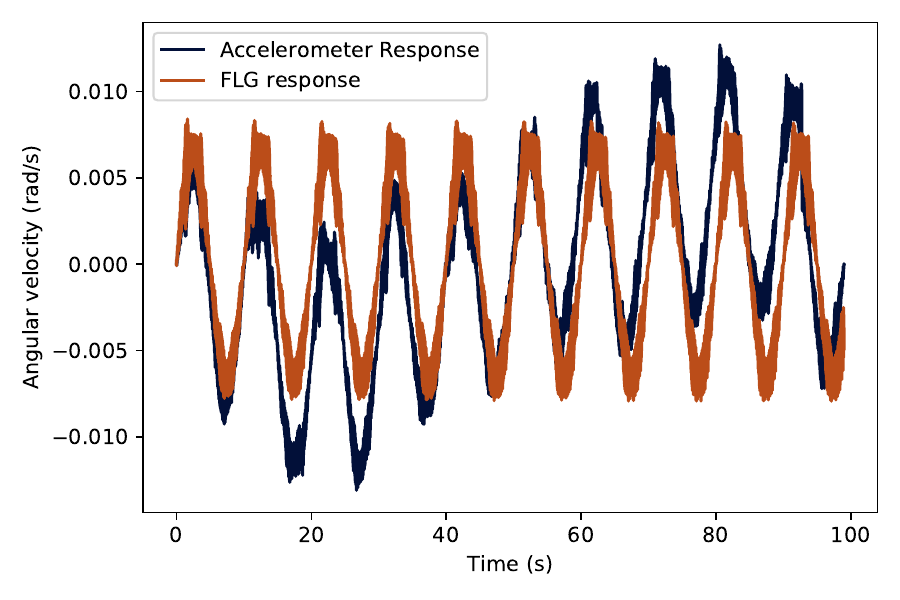}
    \caption{Plot of the fibre laser gyroscope (FLG) output angular velocity against the integrated acceleration measured by accelerometers for a sinusoidal input at 0.1~Hz.}
    \label{fig:gyro_gain}
\end{figure}

Critically, the FLG is used to measure the angle of rotation about the axis orthogonal to the star and baseline vector. The FLG and star tracker together enables the fine metrology measurements to be moved from the chief rigid body frame to an intertial frame, in order to predict open-loop fringe motion. We characterised the FLG to ensure it was within specifications, through driving the gyroscope with a sinusoidal input at 0.1~Hz. A plot of the FLG voltage, converted into angular velocity measurements through a gain of 0.152 rad/s/V, against the integrated accelerations of the accelerometers, again in rad/s, is shown in Figure \ref{fig:gyro_gain}. We see that the FLG does not drift substantially over a long 100~s test. The RMS noise of the FLG was found to be 9.21$\times10^{-6}$ rad/s and the bias was calculated as $5.74\times10^{-6}$ rad/s, indicating that at a power bandwidth of 60~Hz, the FLG exhibits a random walk of 0.26"/s, which is within the desired specification of 0.003 deg/hr$^{1/2}$.

\begin{figure}
    \centering
    \includegraphics[width=\columnwidth]{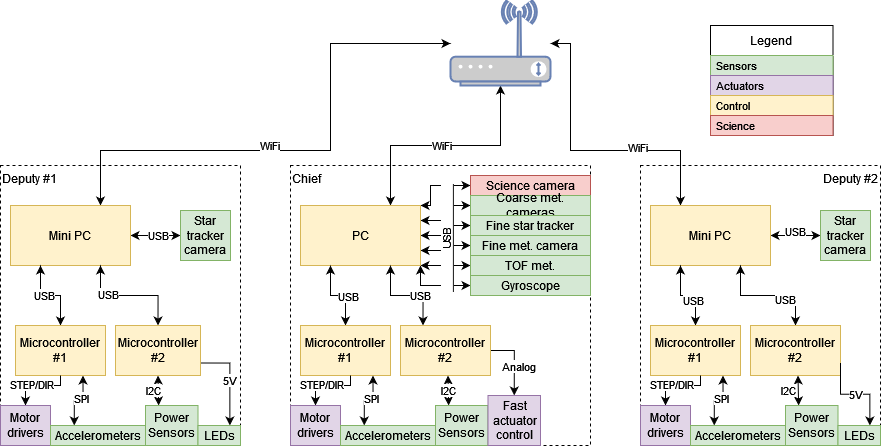}
    \vspace{0.2cm}
    \caption{Physical architecture of control elements and interfaces}
    \label{fig:control_arch}
\end{figure}

% control computer and interfaces (needs diagram!)
Interfacing the array of sensors and actuators is a set of computers comprising the physical elements of the control system. On the chief robot is a PC with a 6-core i5-8500 processor, responsible for tracking the dynamic state of all three robots, as well as coordinating the many systems mounted on the chief robot (fibre optic gyroscope, beam combiner, time-of-flight metrology, etc.). Multiple Teensy 4.1 microcontrollers are also connected to the PC in order to manage all non-USB interfaces. The PC is connected to WiFi, over which requests can be sent to its servers, and over which it can send requests to servers running on the deputy robot computers. The deputy robots mirror this configuration, but with a mini PC (Intel NUC with 4-core i5-10210U processors) reflecting the fewer number of USB connections required and the reduced computational demands. A schematic of this interface architecture for the control system can be found in Figure \ref{fig:control_arch}.

In order to integrate this large number of different physical components, each requiring different communication channels and protocols, a server-based software architecture has been designed and implemented for a number of the systems. In this architecture, each interface is given a server that can respond to a variety of requests depending on the nature of the system it interfaces. For example, a microcontroller on each deputy robot is used to monitor voltage and current draw, as well as control the LEDs used by the coarse camera metrology. This microcontroller connects via USB interface to the control computer, where a server program runs managing it. The server has a set of requests it can respond to by calling various functions, such as reporting the latest voltage measurement, with requests coming either locally from other servers running on the control computer, or over WiFi from the chief or the user interface.

The control system itself is implemented through a PID (proportional–integral–derivative) controller, that aims to minimise the error in linear and angular position and velocity provided from the sensors through a feedback loop. That is, for a given desired state $r(t)$, and a measured state $y(t)$, the control function $u(t)$ in terms of the error value $e(t)$ is:
\begin{align}
    e(t) &= r(t) - y(t)\\
    u(t) &= K_1e(t) + K_2\int^t_0e(x)dx + K_3\dv{e(t)}{t}
\end{align}
where $K_1$, $K_2$ and $K_3$ are tuning variables that are used to maximise the performance of the control loop. A linear-quadratic-regulator (LQG) controller was also considered, but was deemed unnecessary for the types of error correction needed in \textit{Pyxis}.

\subsection{Pointing Control and Star Tracking}
\label{sec:startracker}
To ensure that \textit{Pyxis} has sufficient attitude control, we rely on \red{a star tracker} on each deputy platform to provide an estimate of \red{the orientation} of the robot. These star trackers take an image of the sky using a 5$^{\circ}$ FOV, $f=50$~mm lens at a rate of approximately 3~Hz. This provides a balance between attitude update speed and the ability to detect numerous stars. An algorithm incorporating the Tetra3\footnote{GitHub: \url{https://github.com/esa/tetra3}} and Astrometry.net\cite{Lang2010}\footnote{Website: \url{https://astrometry.net}} plate solvers then extracts the centroid positions of the brightest stars in the image and matches them to a set of known stellar positions located in an index file. These index files were compiled for Astrometry.net utilising the Tycho-2 catalogue\cite{Hog2000}. The matched positions are then used to provide an estimate for the right ascension ($\alpha$) and declination ($\delta$) of the centre of the image, as well as the position angle of the image (angle of the top centre of the image from the North Celestial Pole). These angles are converted into an altitude/azimuth/position angle quaternion for use in correcting and adjusting the attitude of the robot. 

We tested the plate solving algorithm for two quantities: speed and accuracy. The former is particularly important, as it cannot be slower than the frame rate of the camera and, by extension, the attitude update speed. It is for this reason that we adapted numerous plate solving algorithms to create a high-speed version. In our tests, we found that the program could output an attitude quaternion from an image in between 0.2 and 0.3 seconds ($\sim$ 4~Hz), which is sufficient for a frame rate of 3~Hz. We then tested the accuracy by obtaining a number of random on-sky images using the same lens, running the plate solver, and comparing the extracted positions with the positions located in the index file. Each image was manually checked to ensure that the plate solver matched with the correct stars. We converted the ($\alpha,\delta$) coordinates of each star into polar coordinates to give us an estimate of the radial position error and the azimuthal orientation error. We found that the RMS error in the radial direction was about 2", and the RMS error in orientation was 30". Our requirements are that the deputy platforms are able to measure their angle to $\pm$100" in an angle about the star pointing vector (that is, the position angle of the image) and $\pm20$" in the other two axes. Hence, our plate solver will be adequate in accuracy to function as an attitude estimator for \textit{Pyxis}.

The fine star tracker on the chief platform also utilises a plate solving scheme, although it has a much tighter angular position requirement of 0.2" along the optical path axis, and as such utilises a much smaller field of view (approximately 1~degree). This leads to a much slower solve between 1.5 and 2~Hz. To augment this, we implement a centroiding algorithm that augments the star tracker at 3-4~Hz and tracks the position of the target star, using a sufficiently long exposure time so that the star is blurred through seeing (avoiding the noise introduced through the movement of a star's position by the same seeing). Of course, the critical angle of alignment of the chief platform is managed by the FLG, and so the star tracking system acts as a fallback for the rest of the attitude determination.

\subsection{Tip/Tilt Control}
\label{sec:tip_tilt_control}

This tip/tilt control of the starlight injection into the chief platform, described in Section \ref{sec:FI}, is measured though a weighted centre of gravity (WCOG) centroiding algorithm, given by
\begin{align}
    \vb{c} = \frac{\sum_{ij}{w(\vb{x}_{ij})F(\vb{x}_{ij})\vb{x}_{ij}}}{\sum_{ij}{w(\vb{x}_{ij})F(\vb{x}_{ij})}},
\end{align}
where $F(\vb{x_{ij}})$ is the flux of a given pixel $x_{ij}$ and $w(x_{ij})$ is a super-Gaussian weighting function of the form:
\begin{align}
    w(x,y) = e^{-\frac{1}{4\sigma^4}((x-x_0)^2+(y-y_0)^2)^2}.
\end{align}
The centroid is measured with respect to the tip/tilt reference position in pixel coordinates, described in Section \ref{sec:system_alignment}.

Now, the difference between the measured centroid and the reference position is controlled to zero through two control systems with differing timescales. On short timescales (around 200~Hz), the position error is directly sent to the X-Y piezo stages and controlled through a proportional controller. Due to the relatively small stroke of the actuators, however, we then implement a second control loop on longer timescales (about 5~Hz). The positional error is sent to the star tracker on the relevant deputy platform, and converted into a motion to move the deputy angle to alleviate the reliance on the piezos. 

This motion is calculated via applying an offset to the reference pixel of the star tracker. That is, the plate solver solves for the location of the field of view offset to the centre of the image; this is also required for accounting for the offset between the star tracking camera and telescope. To convert between the tip/tilt camera pixel frame ($\vb{x}$) and the star tracker camera frame ($\vb{x'}$), we use the following conversion matrix:
\begin{align}
    \vb{x} = \begin{bmatrix}
    \pm \cos^2{\epsilon} & \pm5 \\
    \cos{\epsilon}\sin{\epsilon} + 5 \cos^2{\epsilon} & 1
\end{bmatrix}\vb{x'},
\end{align}
where $\epsilon$ is the elevation pointing angle of the deputy, the factor of 5 comes from the telescope magnification and the relevant signs are flipped for the two deputies. We also note that before the control loop is closed during alignment, the server controlling the tip/tilt system can still send alignment corrections to the star tracker, so that the centroid spot is located well within the range of the tip/tilt piezos.

\subsection{Fringe Tracking Control}
\label{sec:fringe_tracking_control}
The fringe tracker relies on a group delay estimator of the spectrally dispersed fringes. This estimator comes from the production of an array of $\vb{P2VM}$ matrices for each spectral channel and polarisation, calculated using the output flux ratios of each input following the method outlined in Hansen et. al. (2022)\cite{Hansen2022}. To begin with, a matrix of trial delay phasors, $\vb{\tau}$ were calculated:
\begin{align}
    \vb{x} &= [-a,-a+\delta,...,a-\delta,a]\\
    \vb*{\tau} &= e^{2\pi i \left(\vb{x}\otimes\frac{1}{\vb*{\lambda}}\right)},
\end{align}
where $a$ is half the coherence length, $\delta$ is the group delay resolution, $\vb{x}$ is the vector of trial delays and $\vb*{\lambda}$ is the vector of wavelength channels.

As described in Hansen et. al. (2022)\cite{Hansen2022}, we obtain the complex coherence of each $i$th wavelength channel and polarisation by multiplying the instantaneous intensity by the $\vb{P2VM}$ matrix:
\begin{align}
        \begin{bmatrix}
    \mathfrak{R}(\gamma_i)F_{0,i}\\
    \mathfrak{I}(\gamma_i)F_{0,i}\\
    F_{0,i}
    \end{bmatrix} &= \mathbf{P2VM}_i\cdot\mathbf{I}_i \\
    \gamma_i &= \mathfrak{R}(\gamma_i) + i\mathfrak{I}(\gamma_i).
\end{align}

The fringe group delay envelope, $H$ is simply the multiplication of the complex coherence vector with the trial delay phasor matrix, effectively sampling the Fourier transform of the coherence.
\begin{align}
    H(x) = \vb*{\tau}\cdot\vb*{\gamma}.
\end{align}
The group delay \red{$x_\text{gd}$} is then the delay $x$ corresponding to the maximum of the power spectrum $P(x) = |H(x)|^2$ of this envelope.

To reduce the effect of calibration error in the $\vb{P2VM}$ matrix, we subtract a ``foreground'' power spectrum derived from the intensity of the combined beams a long way away from the fringe envelope. Furthermore, to mitigate some of the effects of scintillation and rapid variance in the group delay estimator, we employ a fading memory controller, where each instantaneous delay is combined with the average of the previous delays $(\bar{P}(x))$ scaled by a factor $\alpha$. Combining these two effects, we obtain:
\begin{align}
    \bar{P}(x)_i = \alpha (P(x)_i - P(x)_\text{foreground}) + (1-\alpha)\bar{P}(x)_{i-1},
\end{align}
where the group delay associated with $\bar{P}(x)_i$ is sent to the controller.

\begin{figure}
    \centering
    \includegraphics[width=\columnwidth]{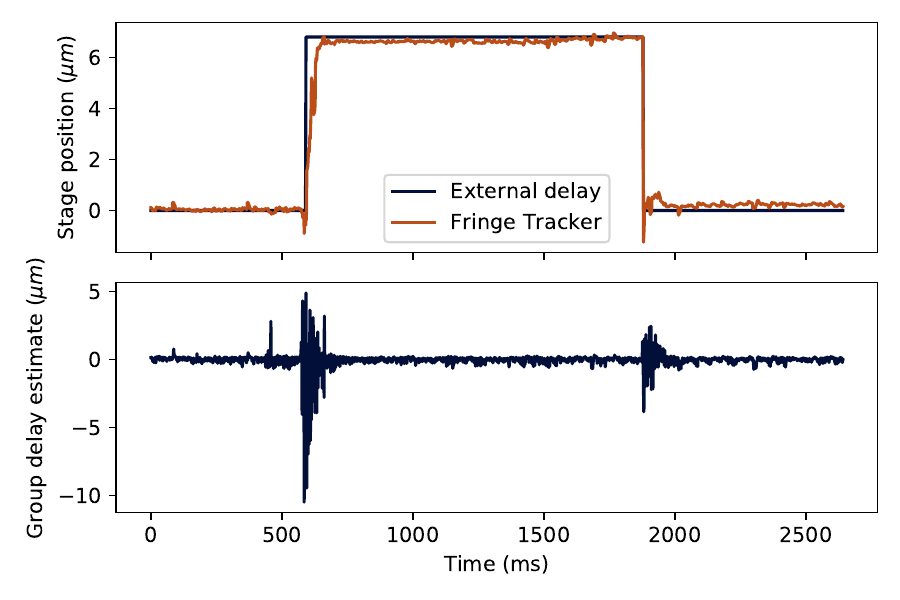}
    \vspace{0.2cm}
    \caption{Response of the fringe tracking servo loop to a commanded top-hat external delay modulation, and associated group delay estimate.}
    \label{fig:fringe_servo}
\end{figure}

The fringe tracker utilises a proportional velocity controller, where the speed of the delay line is proportional to the group delay estimator, scaled by a gain factor $\beta$. Specifically, at each estimation of the group delay, the SmarAct stage is given a command to move a set distance, clocking with steps at a period given by:
\begin{align}
    p = 20\frac{\beta}{x_\text{gd}},
\end{align}
where the prefactor is used to scale the units of group delay into step counts and the gain into units of milliseconds. Each estimation overrides the previous command, providing a smoother control than implementing a positional controller. The performance of the servo loop can be seen in Figure \ref{fig:fringe_servo}, using a gain of $\beta = 50$~ms and a fading memory parameter of $\alpha = 0.95$. The external delay was purposely moved forwards and backwards by approximately 7~\textmu m to be corrected by the fringe tracker. Note that the external delay was not perfectly accurate due to the tolerances on the stage used, and the external stage also exhibited oscillatory behaviour (as shown by the large oscillations of the group delay estimate). Nevertheless, we see that the controller responds well to external delay movements, and holds the group delay constant at zero with an RMS of approximately 200~nm. Further optimisation of the parameters $\alpha$ and $\beta$ will be done on-sky.

The fringe envelope can also be used to perform a signal to noise (SNR) estimation for fringe searching in the case of a failing or absent fine metrology system. The signal is given by the maximum of the foreground subtracted power spectrum, and the background is provided by RMS of the median of the real and imaginary components of the envelope:
\begin{align}
    r &= \text{median}(|\mathfrak{R}(H(x))|^2)\\
    i &= \text{median}(|\mathfrak{I}(H(x))|^2)\\
    \text{SNR} &= \frac{\text{max}(P(x)_i - P(x)_\text{foreground})}{\sqrt{r^2+i^2}}
\end{align}
The SmarAct stage, assuming that the coarse metrology system has equalised the baselines up to the stroke of 8~mm, can then scan for the fringe envelope and stop when the SNR reaches a predetermined threshold.

\subsection{Concept of Operations}

\red{The initial concept of operations for Pyxis includes not only the control loops, but also the steps required to acquire a star, and lock all loops. These are:

\begin{enumerate}
\item Wooden bases for each platform are positioned manually and levelled based on an observation plan. This process takes 2 people and less than 2 minutes per base. This manual positioning is completed to within 20\,cm.
\item The chief and both deputies are commanded to start their star trackers and point at the star. As the star trackers can solve for platform orientation in any pointing direction, this process is only limited by the slew speed of approximately 1 degree per second.
\item The chief coarse metrology system acquires the LEDs on the deputies, moving the deputies into the correct position in 3 dimensions, while the star trackers continue to command the deputy angle.
\item With the deputies at the correct angle to within 30\,arcseconds and the correct position to within 10\,mm, the star from each deputy is within the field of view of the tip/tilt camera. 
\item After the tip/tilt camera is locked, the combination of star trackers, tip/tilt camera and coarse metrology have full control over all axes, with limited precision along the baseline direction.
\item A fringe search is next started, by moving the deputy platforms along the baseline direction. Once fringes are found, this along-baseline motion is controlled by the science camera and fringe tracking stage.
\end{enumerate}

As noted in section 2, this concept is not precisely compatible with space, where thrusters along the baseline direction may not be available. We intend to explore alternative control concepts designed to simulate space once all subsystems and normal operations are fully commissioned.
}

%%%%%%%%%%%%%%%%%%%%%%%%%%%%%%%%%%%%%%%%%%%%%%%%%%%%%%%%%%%%%%%%%%%%%%%%%
\section{Summary}

The \textit{Pyxis} interferometer, when complete, will be a critical step towards verifying the technological readiness of space-based interferometry in the search for Earth-like exoplanets. Specifically, utilising its free-form platform nature and multi-stage metrology system, it will provide a demonstration of satellite-like formation-flying without the cost of space qualification and launch. Furthermore, due to its relatively simple optical design and beam combiner, it will be able to do unique visible-light polarimetric interferometry; the only instrument of its kind in the Southern Hemisphere.

Over the next few years, our goal will be to conduct on-sky demonstrations and observations, while simultaneously preparing for the next phase in the project: a satellite version of the same system (see Hansen and Ireland (2020)\cite{Hansen2020}). To our knowledge, this would be the first demonstration of optical interferometric fringe tracking in space, and furthermore that utilising multiple separate spacecraft in formation. With these demonstrators, it is the authors' hope that the technological barriers will be sufficiently eased such that the final goal of a large scale, mid-infrared space interferometer such as LIFE will be achievable in the coming decades. For it is only such a mission that will truly begin to probe one of the greatest scientific questions of our time: ``are there habitable worlds out there?''

%%%%%%%%%%%%%%%%%%%%%%%%%%%%%%%%%%%%%%%%%%%%%%%%%%%%%%%%%%%%%%%%%%%%%%%%%

%\appendix    % this command starts appendixes

% \disclosures 
\subsection*{Disclosures}
This paper was derived and adapted from two 2022 SPIE Astronomical Telescopes and Instrumentation conference proceedings: Paper 12183-1B ``The Pyxis Interferometer (I): Scientific Context, Metrology System and Optical Design''\cite{Hansen2022Spie} and Paper 12183-1C ``The Pyxis Interferometer (II): Control System, Telescope and Mechanical Design'' \cite{Wade2022}. The fine metrology section (Section \ref{sec:fine_metrology}) also adapts a portion of the 2020 SPIE Astronomical Telescopes and Instrumentation conference proceeding paper 11446-2F "Compact unambiguous differential path-length metrology with dispersed Fabry-Perot laser diodes for a space interferometer array"\cite{Lagadec2020}.

\subsection* {Code, Data, and Materials Availability} 
The data used in this paper is available upon reasonable request to the authors.

\subsection* {Acknowledgments}
We acknowledge and celebrate the traditional custodians of the land on which the Australian National University is based, the Ngunnawal and Ngambri peoples, and pay our respects to elders past and present.

The authors also acknowledge the substantial work of the entire \textit{Pyxis} team and associates in progressing this project and its results: Julien Bernard, Nicholas Bohlsen, Logan Corry, Michael Ellis, Steven Ellis, Alex Fan, Shanae King, Weihao Luo, Stephen Madden, Joseph Mangos, Patrick Miller, Michael Polkinghorne, Laura Schlueter, Thomas Scott, Hancheng Shao and Kunlun Yan.

This research was supported by funding from Australian Research Council grant No. DP200102383. JH acknowledges support from the Australian Government Research Training Program, and the College of Science's Dean's Merit HDR supplementary scholarship.

%%%%% References %%%%%

\bibliography{article}   % bibliography data in report.bib

\begin{thebibliography}{10}

\bibitem{Abuter2017}
R.~Abuter, M.~Accardo, A.~Amorim, {\em et~al.}, ``First light for {GRAVITY}:
  {Phase} referencing optical interferometry for the {Very} {Large} {Telescope}
  {Interferometer},'' {\em Astronomy \& Astrophysics} {\bf 602}, A94  (2017).
\newblock Publisher: EDP Sciences.

\bibitem{Lopez2022}
B.~{Lopez}, S.~{Lagarde}, R.~G. {Petrov}, {\em et~al.}, ``{MATISSE, the VLTI
  mid-infrared imaging spectro-interferometer},'' {\em Astronomy \&
  Astrophysics} {\bf 659}, A192  (2022).

\bibitem{Anugu2020}
N.~{Anugu}, J.-B. {Le Bouquin}, J.~D. {Monnier}, {\em et~al.}, ``{MIRC-X: A
  Highly Sensitive Six-telescope Interferometric Imager at the CHARA Array},''
  {\em The Astronomical Journal} {\bf 160}, 158  (2020).

\bibitem{Roettenbacher2016}
R.~M. {Roettenbacher}, J.~D. {Monnier}, H.~{Korhonen}, {\em et~al.}, ``{No
  Sun-like dynamo on the active star {\ensuremath{\zeta}} Andromedae from
  starspot asymmetry},'' {\em Nature} {\bf 533}, 217--220  (2016).

\bibitem{Lacour2021}
S.~{Lacour}, J.~J. {Wang}, L.~{Rodet}, {\em et~al.}, ``{The mass of
  {\ensuremath{\beta}} Pictoris c from {\ensuremath{\beta}} Pictoris b orbital
  motion},'' {\em Astronomy \& Astrophysics} {\bf 654}, L2  (2021).

\bibitem{Gravity2019}
{Gravity Collaboration}, R.~{Abuter}, A.~{Amorim}, {\em et~al.}, ``{A geometric
  distance measurement to the Galactic center black hole with 0.3\%
  uncertainty},'' {\em Astronomy \& Astrophysics} {\bf 625}, L10  (2019).

\bibitem{Schwieterman2018}
E.~W. {Schwieterman}, N.~Y. {Kiang}, M.~N. {Parenteau}, {\em et~al.},
  ``{Exoplanet Biosignatures: A Review of Remotely Detectable Signs of Life},''
  {\em Astrobiology} {\bf 18}, 663--708  (2018).

\bibitem{DiamondLowe2020}
H.~{Diamond-Lowe}, Z.~{Berta-Thompson}, D.~{Charbonneau}, {\em et~al.},
  ``{Simultaneous Optical Transmission Spectroscopy of a Terrestrial,
  Habitable-zone Exoplanet with Two Ground-based Multiobject Spectrographs},''
  {\em The Astronomical Journal} {\bf 160}, 27  (2020).

\bibitem{Cockell2009}
C.~S. {Cockell}, T.~{Herbst}, A.~{L{\'e}ger}, {\em et~al.},
  ``{Darwin{\textemdash}an experimental astronomy mission to search for
  extrasolar planets},'' {\em Experimental Astronomy} {\bf 23}, 435--461
  (2009).

\bibitem{Defrere2018}
D.~{Defr{\`e}re}, A.~{L{\'e}ger}, O.~{Absil}, {\em et~al.}, ``{Space-based
  infrared interferometry to study exoplanetary atmospheres},'' {\em
  Experimental Astronomy} {\bf 46}, 543--560  (2018).

\bibitem{Kammerer2018}
J.~{Kammerer} and S.~P. {Quanz}, ``{Simulating the exoplanet yield of a
  space-based mid-infrared interferometer based on Kepler statistics},'' {\em
  Astronomy \& Astrophysics} {\bf 609}, A4  (2018).

\bibitem{LIFE1}
S.~P. {Quanz}, M.~{Ottiger}, E.~{Fontanet}, {\em et~al.}, ``{Large
  Interferometer For Exoplanets (LIFE). I. Improved exoplanet detection yield
  estimates for a large mid-infrared space-interferometer mission},'' {\em
  Astronomy \& Astrophysics} {\bf 664}, A21  (2022).

\bibitem{Leisawitz2007}
D.~{Leisawitz}, C.~{Baker}, A.~{Barger}, {\em et~al.}, ``{The space infrared
  interferometric telescope (SPIRIT): High-resolution imaging and spectroscopy
  in the far-infrared},'' {\em Advances in Space Research} {\bf 40}, 689--703
  (2007).

\bibitem{Unwin2008}
S.~C. {Unwin}, M.~{Shao}, A.~M. {Tanner}, {\em et~al.}, ``{Taking the Measure
  of the Universe: Precision Astrometry with SIM PlanetQuest},'' {\em
  Publications of the Astronomical Society of the Pacific} {\bf 120}, 38
  (2008).

\bibitem{LeDuigou2006}
J.~M. {Le Duigou}, M.~{Ollivier}, A.~{L{\'e}ger}, {\em et~al.}, ``{Pegase: a
  space-based nulling interferometer},'' in {\em Society of Photo-Optical
  Instrumentation Engineers (SPIE) Conference Series},  J.~C. {Mather}, H.~A.
  {MacEwen}, and M.~W.~M. {de Graauw}, Eds., {\em Society of Photo-Optical
  Instrumentation Engineers (SPIE) Conference Series} {\bf 6265}, 62651M
  (2006).

\bibitem{Beichman1999}
C.~A. Beichman, N.~J. Woolf, and C.~A. Lindensmith, ``{The Terrestrial Planet
  Finder (TPF) : a NASA Origins Program to search for habitable planets},''
  tech. rep., Jet Propulsion Laboratory, California Institute of Technology,
  Pasadena, California  (1999).

\bibitem{Borucki2010}
W.~J. {Borucki}, D.~{Koch}, G.~{Basri}, {\em et~al.}, ``{Kepler
  Planet-Detection Mission: Introduction and First Results},'' {\em Science}
  {\bf 327}, 977  (2010).

\bibitem{Ricker2015}
G.~R. {Ricker}, J.~N. {Winn}, R.~{Vanderspek}, {\em et~al.}, ``{Transiting
  Exoplanet Survey Satellite (TESS)},'' {\em Journal of Astronomical
  Telescopes, Instruments, and Systems} {\bf 1}, 014003  (2015).

\bibitem{Broeg2013}
C.~{Broeg}, A.~{Fortier}, D.~{Ehrenreich}, {\em et~al.}, ``{CHEOPS: A transit
  photometry mission for ESA's small mission programme},'' in {\em European
  Physical Journal Web of Conferences},  {\em European Physical Journal Web of
  Conferences} {\bf 47}, 03005  (2013).

\bibitem{LIFE2}
F.~A. {Dannert}, M.~{Ottiger}, S.~P. {Quanz}, {\em et~al.}, ``{Large
  Interferometer For Exoplanets (LIFE). II. Signal simulation, signal
  extraction, and fundamental exoplanet parameters from single-epoch
  observations},'' {\em Astronomy \& Astrophysics} {\bf 664}, A22  (2022).

\bibitem{ESAVoyage2021}
{Voyage 2050 Senior Committee}, {\em Voyage 2050 - Final Recommendations from
  the Voyage 2050 Senior Committee}, European Space Agency  (2021).

\bibitem{Ranganathan2022}
M.~{Ranganathan}, A.~M. {Glauser}, T.~{Birbacher}, {\em et~al.}, ``{The Nulling
  Interferometer Cryogenic Experiment: I},'' in {\em Optical and Infrared
  Interferometry and Imaging VIII},  A.~{M{\'e}rand}, S.~{Sallum}, and
  J.~{Sanchez-Bermudez}, Eds., {\em Society of Photo-Optical Instrumentation
  Engineers (SPIE) Conference Series} {\bf 12183}, 121830L  (2022).

\bibitem{Dandumont2020}
C.~{Dandumont}, D.~{Defr{\`e}re}, J.~{Kammerer}, {\em et~al.}, ``{Exoplanet
  detection yield of a space-based Bracewell interferometer from small to
  medium satellites},'' {\em Journal of Astronomical Telescopes, Instruments,
  and Systems} {\bf 6}, 035004  (2020).

\bibitem{Matsuo2022}
T.~{Matsuo}, S.~{Ikari}, H.~{Kondo}, {\em et~al.}, ``{High spatial resolution
  spectral imaging method for space interferometers and its application to
  formation flying small satellites},'' {\em Journal of Astronomical
  Telescopes, Instruments, and Systems} {\bf 8}, 015001  (2022).

\bibitem{Hansen2020}
J.~T. {Hansen} and M.~J. {Ireland}, ``{A linear formation-flying astronomical
  interferometer in low Earth orbit},'' {\em Publications of the Astronomical
  Society of Australia} {\bf 37}, e019  (2020).

\bibitem{Davis1999}
J.~{Davis}, W.~J. {Tango}, A.~J. {Booth}, {\em et~al.}, ``{The Sydney
  University Stellar Interferometer - I. The instrument},'' {\em Monthly
  Notices of the Royal Astronomical Society} {\bf 303}, 773--782  (1999).

\bibitem{Armstrong1998}
J.~T. Armstrong, D.~Mozurkewich, L.~J. Rickard, {\em et~al.}, ``The {Navy}
  {Prototype} {Optical} {Interferometer},'' {\em The Astrophysical Journal}
  {\bf 496}, 550  (1998).
\newblock Publisher: IOP Publishing.

\bibitem{Tuthill2000}
P.~G. {Tuthill}, J.~D. {Monnier}, W.~C. {Danchi}, {\em et~al.}, ``{Michelson
  Interferometry with the Keck I Telescope},'' {\em Publications of the
  Astronomical Society of the Pacific} {\bf 112}(770), 555--565  (2000).

\bibitem{Clark2021}
J.~T. {Clark}, M.~{Clert{\'e}}, N.~R. {Hinkel}, {\em et~al.}, ``{The GALAH
  Survey: using galactic archaeology to refine our knowledge of TESS target
  stars},'' {\em Monthly Notices of the Royal Astronomical Society} {\bf 504},
  4968--4989  (2021).

\bibitem{Rains2021}
A.~D. {Rains}, M.~{{\v{Z}}erjal}, M.~J. {Ireland}, {\em et~al.},
  ``{Characterization of 92 southern TESS candidate planet hosts and a new
  photometric [Fe/H] relation for cool dwarfs},'' {\em Monthly Notices of the
  Royal Astronomical Society} {\bf 504}, 5788--5805  (2021).

\bibitem{Tayar2022}
J.~{Tayar}, Z.~R. {Claytor}, D.~{Huber}, {\em et~al.}, ``{A Guide to Realistic
  Uncertainties on the Fundamental Properties of Solar-type Exoplanet Host
  Stars},'' {\em The Astrophysical Journal} {\bf 927}, 31  (2022).

\bibitem{White2013}
T.~R. {White}, D.~{Huber}, V.~{Maestro}, {\em et~al.}, ``{Interferometric radii
  of bright Kepler stars with the CHARA Array: {\ensuremath{\theta}} Cygni and
  16 Cygni A and B},'' {\em Monthly Notices of the Royal Astronomical Society}
  {\bf 433}, 1262--1270  (2013).

\bibitem{Epstein2014}
C.~R. {Epstein}, Y.~P. {Elsworth}, J.~A. {Johnson}, {\em et~al.}, ``{Testing
  the Asteroseismic Mass Scale Using Metal-poor Stars Characterized with APOGEE
  and Kepler},'' {\em The Astrophysical Journal Letters} {\bf 785}, L28
  (2014).

\bibitem{Valentini2019}
M.~{Valentini}, C.~{Chiappini}, D.~{Bossini}, {\em et~al.}, ``{Masses and ages
  for metal-poor stars. A pilot programme combining asteroseismology and
  high-resolution spectroscopic follow-up of RAVE halo stars},'' {\em Astronomy
  and Astrophysics} {\bf 627}, A173  (2019).

\bibitem{Huber2012}
D.~{Huber}, M.~J. {Ireland}, T.~R. {Bedding}, {\em et~al.}, ``Fundamental
  properties of stars using asteroseismology from kepler and corot and
  interferometry from the chara array,'' {\em The Astrophysical Journal} {\bf
  760}, 32  (2012).

\bibitem{Ireland2005}
M.~J. {Ireland}, P.~G. {Tuthill}, J.~{Davis}, {\em et~al.}, ``{Dust scattering
  in the Miras R Car and RR Sco resolved by optical interferometric
  polarimetry},'' {\em Monthly Notices of the Royal Astronomical Society} {\bf
  361}, 337--344  (2005).

\bibitem{Norris2012}
B.~R.~M. {Norris}, P.~G. {Tuthill}, M.~J. {Ireland}, {\em et~al.}, ``{A close
  halo of large transparent grains around extreme red giant stars},'' {\em
  Nature} {\bf 484}, 220--222  (2012).

\bibitem{Hoefner2018}
S.~{H{\"o}fner} and H.~{Olofsson}, ``{Mass loss of stars on the asymptotic
  giant branch. Mechanisms, models and measurements},'' {\em The Astronomy and
  Astrophysics Review} {\bf 26}, 1  (2018).

\bibitem{Lagadec2020}
T.~{Lagadec}, M.~{Ireland}, J.~{Hansen}, {\em et~al.}, ``{Compact unambiguous
  differential path-length metrology with dispersed Fabry-Perot laser diodes
  for a space interferometer array},'' in {\em Society of Photo-Optical
  Instrumentation Engineers (SPIE) Conference Series},  {\em Society of
  Photo-Optical Instrumentation Engineers (SPIE) Conference Series} {\bf
  11446}, 114462F  (2020).

\bibitem{Polhemus1973}
C.~Polhemus, ``Two-wavelength interferometry,'' {\em Appl. Opt.} {\bf 12},
  2071--2074  (1973).

\bibitem{Osellame2012}
R.~{Osellame}, G.~{Cerullo}, and R.~{Ramponi}, {\em {Femtosecond Laser
  Micromachining}}, vol.~123, Springer  (2012).

\bibitem{Gross2015}
S.~{Gross} and M.~J. {Withford}, ``{Ultrafast-laser-inscribed 3D integrated
  photonics: challenges and emerging applications},'' {\em Nanophotonics} {\bf
  4}, 20  (2015).

\bibitem{Norris2020}
B.~R.~M. {Norris}, N.~{Cvetojevic}, T.~{Lagadec}, {\em et~al.}, ``{First on-sky
  demonstration of an integrated-photonic nulling interferometer: the GLINT
  instrument},'' {\em Monthly Notices of the Royal Astronomical Society} {\bf
  491}, 4180--4193  (2020).

\bibitem{Hansen2022}
J.~T. Hansen, M.~J. Ireland, A.~Ross-Adams, {\em et~al.}, ``{Interferometric
  beam combination with a triangular tricoupler photonic chip},'' {\em Journal
  of Astronomical Telescopes, Instruments, and Systems} {\bf 8}(2), 025002
  (2022).

\bibitem{Lang2010}
D.~{Lang}, D.~W. {Hogg}, K.~{Mierle}, {\em et~al.}, ``{Astrometry.net: Blind
  Astrometric Calibration of Arbitrary Astronomical Images},'' {\em The
  Astronomical Journal} {\bf 139}, 1782--1800  (2010).

\bibitem{Hog2000}
E.~{H{\o}g}, C.~{Fabricius}, V.~V. {Makarov}, {\em et~al.}, ``{The Tycho-2
  catalogue of the 2.5 million brightest stars},'' {\em Astronomy and
  Astrophysics} {\bf 355}, L27--L30  (2000).

\bibitem{Hansen2022Spie}
J.~T. {Hansen}, M.~J. {Ireland}, T.~{Travouillon}, {\em et~al.}, ``{The Pyxis
  Interferometer (I): scientific context, metrology system, and optical
  design},'' in {\em Optical and Infrared Interferometry and Imaging VIII},
  A.~{M{\'e}rand}, S.~{Sallum}, and J.~{Sanchez-Bermudez}, Eds., {\em Society
  of Photo-Optical Instrumentation Engineers (SPIE) Conference Series} {\bf
  12183}, 121831B  (2022).

\bibitem{Wade2022}
S.~{Wade}, J.~T. {Hansen}, M.~J. {Ireland}, {\em et~al.}, ``{The Pyxis
  Interferometer (II): control system, telescope, and mechanical design},'' in
  {\em Optical and Infrared Interferometry and Imaging VIII},  A.~{M{\'e}rand},
  S.~{Sallum}, and J.~{Sanchez-Bermudez}, Eds., {\em Society of Photo-Optical
  Instrumentation Engineers (SPIE) Conference Series} {\bf 12183}, 121831C
  (2022).

\end{thebibliography}
\bibliographystyle{spiejour}   % makes bibtex use spiejour.bst

%%%%% Biographies of authors %%%%%

\vspace{2ex}\noindent\textbf{Jonah Hansen} is a PhD candidate at the Australian National University, researching into space interferometry. In particular, he is assisting to design and build the \textit{Pyxis} interferometer - a ground based pathfinder for an eventual space interferometer mission, LIFE (Large Interferometer For Exoplanets), designed to find and characterise exoplanets. He is also assisting in modelling different array architectures and beam combiners for this latter mission. 

\vspace{2ex}\noindent\textbf{Samuel Wade} is a mechatronics and project engineer at the Australian National University, assisting in the engineering aspects of \textit{Pyxis} - a ground-based pathfinder for an optical space interferometer. Samuel has had previous experience in satellite sensors and ground-based space surveillance.

\vspace{2ex}\noindent\textbf{Michael Ireland} is a Professor of Astrophysics and Instrumentation Science at the Australian National University. He obtained his PhD from the University of Sydney in 2006, and has since held positions at the California Institute of Technology, the Australian Astronomical Observatory, Macquarie University and the University of Sydney. His research focuses on stellar astrophysics, exoplanet formation, the search for life on other worlds and technologies needed to support these endeavours

\vspace{2ex}\noindent\textbf{Tony Travouillon} is an associate professor at the Australian National University and is responsible for the instrumentation and telescope development of its school of astronomy. He received his PhD in 2005 from the University of New South Wales.

\vspace{2ex}\noindent\textbf{Tiphaine Lagadec} is an instrument scientist whose interest is the development of interferometric techniques and technologies to investigate stellar systems at high resolution. Tiphaine obtained a PhD from the University of Sydney in 2020 and has participated in the fields of intensity interferometry, nulling interferometry, space interferometry and solar coronagraphy. 

\vspace{2ex}\noindent\textbf{Nicholas Herrald} is an opto-mechanical engineer with 12 years experience, working for Research School of Astrophysics and Astronomy at the Australian National University.

\vspace{2ex}\noindent\textbf{Joice Mathew} is working as an instrumentation scientist at the Australian National University (ANU). His research interests include electro-optical payload development, instrument modelling, systems engineering, space instrumentation, and qualification. Joice obtained his Ph.D. in astronomical space instrumentation from the Indian Institute of Astrophysics, Bangalore. Before joining ANU, he worked as a visiting instrument scientist at the Max Planck Institute for Solar System Research, Germany on the Solar Orbiter mission.

\vspace{2ex}\noindent\textbf{Stephanie Monty} is a research associate at the University of Cambridge. She completed her PhD at the Australian National University in 2022. Her research focuses are Galactic Archaeology, stellar dynamics, fibre fed spectrographs and adaptive optics.

\vspace{2ex}\noindent\textbf{Adam Rains} completed his PhD in astronomy and astrophysics at the Australian National University in 2021 and is now a postdoctoral researcher at Uppsala University in Sweden. His research interests sit at the intersection of stellar and exoplanetary astrophysics -- in particular the spectroscopic characterisation of low-mass stars and their planets.

%\listoffigures
%\listoftables

%\end{spacing}
\end{document}